\documentclass[aps,prd,preprint,superscriptaddress,preprintnumbers,longbibliography,floatfix]{revtex4-1}
\usepackage{graphicx,float,wrapfig,subfigure}
\usepackage{amsfonts,amsmath,amssymb,amstext}
\usepackage{latexsym}
\usepackage{bm}
\usepackage{color}
\usepackage[normalem]{ulem}

\newcommand{\be}{\begin{equation}}
\newcommand{\ee}{\end{equation}}
\newcommand{\ba}{\begin{eqnarray}}
\newcommand{\ea}{\end{eqnarray}}

\definecolor{red}{rgb}{0.7,0,0}
\definecolor{green}{rgb}{0,0.5,0}

	\newcommand*{\nm}{\nonumber}

\begin{document}

\title{Exploring $U_A(1)$ symmetry restoration in a magnetic field by meson screening masses in the
NJL model}

\author{Bing-Kai Sheng}
\affiliation{ College of Physics, Jilin University, Changchun 130012, China}
\affiliation{School of Physical Science and Technology, Inner Mongolia University, Hohhot, 010021, China}

\author{Zigeng Ding}
\affiliation{Center for Fundamental Physics, School of Mechanics and Optoelectronic Physics, Anhui University of Science and Technology,
Huainan, Anhui 232001, China}

\author{Danning Li}
\email{lidanning@jnu.edu.cn}
\affiliation{Department of Physics and Siyuan Laboratory, Jinan University, Guangzhou 510632, China}

\author{Xinyang Wang}
\email{wangxy@aust.edu.cn}
\affiliation{Center for Fundamental Physics, School of Mechanics and Optoelectronic Physics, Anhui University of Science and Technology,
Huainan, Anhui 232001, China}

\author{Lang Yu}
\email{yulang@jlu.edu.cn}
\thanks{corresponding author}
\affiliation{ College of Physics, Jilin University, Changchun 130012, China}

\begin{abstract}
We investigate the effective axial $U_A(1)$ symmetry restoration at finite temperature in an external magnetic field using meson screening masses, as well as meson susceptibilities, in a $(2+1)$-flavor Nambu-Jona-Lasinio (NJL) model. To incorporate inverse magnetic catalysis, we employ a magnetic-field-dependent four-quark coupling constrained by the lattice QCD chiral pseudo-critical temperature, with the 't Hooft coupling fixed. In the lattice-improved NJL model, the longitudinal screening mass difference between the $U_A(1)$ partners $\delta^0$ and $\pi^0$ and their normalized meson susceptibility difference exhibit axial magnetic catalysis at low temperatures and axial inverse magnetic catalysis near the chiral crossover, whereas the transverse screening mass difference increases with $eB$ at all temperatures. Despite these different responses, the axial pseudo-critical temperatures extracted from the thermal inflection points of these three alternative order parameters decrease with $eB$, implying that a decreasing pseudo-critical temperature can coexist with an increasing transverse screening mass difference at fixed temperature.
\end{abstract}

\maketitle

\section{Introduction}

The effects of magnetic fields on strongly interacting matter have been extensively studied over the past few decades (see Refs.~\cite{Kharzeev:2015kna,Miransky:2015ava,Andersen:2014xxa,Zhai:2026cud} for recent reviews), since strong magnetic fields are expected to exist in the early Universe~\cite{Vachaspati:1991nm,Enqvist:1993np}, in compact stars such as magnetars~\cite{Duncan:1992hi}, and in noncentral heavy-ion collisions~\cite{Skokov:2009qp,Voronyuk:2011jd,Bzdak:2011yy,Deng:2012pc}. The interplay between magnetic fields and the nonperturbative properties of quantum chromodynamics (QCD) gives rise to a variety of intriguing phenomena, such as the chiral magnetic effect (CME)~\cite{Kharzeev:2007tn,Okorokov:2009bf,Fukushima:2008xe}, magnetic catalysis (MC)~\cite{Klevansky:1989vi,Klimenko:1990rh,Gusynin:1995nb,Shovkovy:2012zn}, inverse magnetic catalysis (IMC)~\cite{Bali:2011qj,Bali:2012zg}, and vacuum superconductivity~\cite{Chernodub:2010qx,Chernodub:2011mc}.

In particular, the breaking and restoration of chiral symmetry, characterized by the chiral condensate, is one of the most important aspects of QCD. Early lattice QCD (LQCD) simulations~\cite{Buividovich:2008wf,Braguta:2010ej,DElia:2010abb,DElia:2011koc,Ilgenfritz:2012fw}, as well as almost all low-energy effective theories and models of QCD~\cite{Klevansky:1989vi,Klimenko:1990rh,Gusynin:1995nb,Shovkovy:2012zn,Shushpanov:1997sf,Agasian:1999sx,Alexandre:2000yf,Agasian:2001hv,Cohen:2007bt,Gatto:2010qs,Gatto:2010pt,Mizher:2010zb,Kashiwa:2011js,Avancini:2012ee,Andersen:2012dz,Scherer:2012nn}, show that the chiral condensate increases as the magnetic field grows, a phenomenon known as magnetic catalysis~\cite{Klevansky:1989vi,Klimenko:1990rh,Gusynin:1995nb,Shovkovy:2012zn}, and that the corresponding chiral pseudo-critical temperature $T_{\mathrm{pc}}$ increases with the magnetic field. However, LQCD simulations at physical quark masses show that, although magnetic catalysis
persists at low temperatures, the condensate is suppressed by the magnetic field near the chiral crossover, a phenomenon known
as inverse magnetic catalysis~\cite{Bali:2011qj,Bali:2012zg}. These simulations also find that the chiral pseudo-critical
temperature decreases with increasing magnetic-field strength~\cite{Bali:2011qj,Bali:2012zg}.

Numerous studies~\cite{Fukushima:2012kc,Kojo:2012js,Bruckmann:2013oba,Chao:2013qpa,Fraga:2013ova,Ferreira:2014kpa,Farias:2014eca,
Yu:2014sla,Andersen:2014oaa,Ferrer:2014qka,Providencia:2014txa,Farias:2016gmy,Mao:2016fha,Mamo:2015dea,Endrodi:2019whh,
Endrodi:2019zrl,Tomiya:2019nym,Li:2016gfn,Rodrigues:2018pep,He:2020fdi} have been carried out to understand the underlying mechanism of this puzzle. Lattice simulations in Ref.~\cite{Bruckmann:2013oba} established that inverse magnetic catalysis is driven by sea-quark effects arising from the rearrangement of gluonic configurations induced by the magnetic field. It is well known that, in the standard NJL (SNJL) model, the constant four-quark coupling $G$ originates from gluon-mediated interactions, but the model exhibits only MC at all temperatures, with the chiral pseudo-critical temperature increasing with the magnetic field. Phenomenologically, one possible approach is to incorporate magnetic-field-dependent or temperature- and magnetic-field-dependent coupling constants into NJL-type models~\cite{Ferreira:2014kpa,Farias:2014eca,Providencia:2014txa,
Farias:2016gmy,Endrodi:2019whh}, which can qualitatively reproduce the IMC effect at high temperatures and the decrease of $T_{\mathrm{pc}}$ with $eB$. This can therefore be regarded as an indirect way of introducing sea-quark effects into effective model descriptions.

In addition to the chiral condensate, meson screening masses in hot and/or magnetized QCD matter have also been extensively studied~\cite{Kunihiro:1991hp,Florkowski:1993br,Ishii:2013kaa,Wang:2013wk,Ishii:2015ira,Cao:2021tcr,Cheng:2010fe,
Maezawa:2013nxa,Kaczmarek:2013kva,Bazavov:2019www,Fayazbakhsh:2012vr,Fayazbakhsh:2013cha,Wang:2017vtn,Sheng:2020hge,
Sheng:2021evj,Ding:2022tqn,Coppola:2024uvz,Ding:2025pbu,Ahmed:2026qzw}, as they can serve as useful probes of the chiral and axial $U_A(1)$ symmetry restorations in QCD. In our previous work~\cite{Sheng:2021evj}, we investigated the chiral phase transition using a two-flavor lattice-improved NJL (LNJL) model with a magnetic-field-dependent coupling constant, by analyzing the longitudinal and transverse screening mass differences between the chiral partners $\pi^0$ and $\sigma$ as effective order parameters. Both differences decrease with increasing $eB$ at high temperatures, and the chiral pseudo-critical temperatures extracted from these differences decrease with increasing $eB$, in qualitative agreement with LQCD results inferred from the quark condensate.

Recently, the influence of magnetic fields on effective axial
$U_A(1)$ symmetry restoration has attracted increasing attention. Although the gluonic
anomaly remains in the divergence of the singlet axial current,
an external magnetic field can modify the gauge-field ensemble
through sea-quark effects, thereby affecting topological
fluctuations and the infrared Dirac spectrum. The resulting
changes in effective $U_A(1)$ symmetry breaking can be probed
through differences between correlation functions of $U_A(1)$
partners. A study within a $(2+1)$-flavor NJL model incorporating
quark anomalous magnetic moments examined the meson susceptibility
difference between the $U_A(1)$ partners $\pi^0$ and
$\delta^0$~\cite{Wang:2021dcy}. It reported an enhancement of this difference
at low temperatures and a suppression at high temperatures,
referred to as axial magnetic catalysis (AMC) and axial inverse
magnetic catalysis (AIMC), respectively, in analogy with their
chiral counterparts. The same study also found that the axial
pseudo-critical temperature, defined by the thermal inflection
point of the meson susceptibility difference, decreases with increasing
magnetic field. However, for certain choices of the quark
anomalous magnetic moments, NJL-model calculations predict a
change from a chiral crossover to a first-order transition at
$eB<1.0~\mathrm{GeV}^2$~\cite{Mei:2020jzn}, in contrast to the crossover
behavior observed in lattice calculations over the corresponding
field range. More recently, LQCD calculations~\cite{Ding:2026ewc} have also reported
AMC at low temperatures and AIMC at sufficiently large $eB$ near the crossover,
as indicated by the same meson susceptibility difference.

In this paper, we extend the two-flavor lattice-improved NJL
model of Ref.~\cite{Sheng:2021evj} to the $(2+1)$-flavor case, and investigate
the axial $U_A(1)$ symmetry restoration in an external
magnetic field. We examine the longitudinal and transverse
screening mass differences between the $U_A(1)$ partners
$\pi^0$ and $\delta^0$, together with their meson susceptibility
difference, and compare the pseudo-critical temperatures
extracted from their thermal inflection points. The model
contains the four-quark coupling $G$
and the six-quark coupling $K$. To incorporate IMC, we follow the prescription of
Ref.~\cite{Ferreira:2014kpa} and determine $G(B)$ by fitting the
magnetic-field dependence of the normalized chiral
pseudo-critical temperature from LQCD~\cite{Bali:2011qj}, while keeping
$K$ fixed.

Note that alternative prescriptions can either determine $G(B)$ by matching the average constituent quark mass~\cite{Endrodi:2019whh,Sheng:2021evj}, $M=(M_u+M_d)/2$, while keeping $K$ fixed, or determine both $G(B)$ and $K(B)$ by matching the constituent quark masses $M_d$ and $M_s$~\cite{Moreira:2020wau}, as inferred from LQCD baryon masses~\cite{Endrodi:2019whh}. A detailed comparison of these two prescriptions will be
presented in a separate work~\cite{Yu:inpreparation}: the former may fail to reproduce the low-temperature
magnetic enhancement of the meson susceptibilities for the chiral partners $\pi^0$ and $\sigma$
observed in LQCD simulations~\cite{Ding:2026ewc}; the latter predicts a change from a chiral crossover
to a first-order transition at $eB\gtrsim 0.6~\mathrm{GeV}^2$, in contradiction with LQCD results~\cite{Bali:2012zg}.
Therefore, neither of these two prescriptions is adopted in the present work.

This paper is organized as follows. In Sec.~II, we extend the two-flavor NJL model to a $(2+1)$-flavor one in the presence of a magnetic field, and describe the method for calculating meson screening masses and meson susceptibilities. In Sec.~III, we present and discuss the numerical results. Finally, we summarize and conclude in Sec.~IV.

\section{formalism}

\subsection{NJL Model}

The Lagrangian density of the $(2+1)$-flavor NJL model in an external magnetic field is given by
\begin{equation}\label{eq:NJL_L}
\begin{aligned}
\mathcal{L}
&= \bar{\psi}\left(i\not{\!\!D}-\hat{m}\right)\psi
+ G \sum_{a=0}^{8}\left[
\left(\bar{\psi}\lambda^a\psi\right)^2
+ \left(\bar{\psi} i\gamma_5\lambda^a\psi\right)^2
\right]\\
&\quad- K\left[
\det\left[\bar{\psi}(1+\gamma_5)\psi\right]
+ \det\left[\bar{\psi}(1-\gamma_5)\psi\right]
\right],
\end{aligned}
\end{equation}
where $\psi=(u,d,s)^T$ is the quark flavor triplet and
$\hat{m}=\mathrm{diag}(m_u,m_d,m_s)$ is the corresponding current-quark mass matrix.
The covariant derivative, $D_{\mu}=\partial_{\mu}+i\hat{Q}eA_{\mu}^{\mathrm{ext}}$,
couples quarks to an external magnetic field $\mathbf{B}=(0,0,B)$ along the positive $z$ direction
through a background Abelian gauge field $A_{\mu}^{\mathrm{ext}}=(0,0,-Bx,0)$.
Here $\hat{Q}=\mathrm{diag}(Q_u,Q_d,Q_s)=\mathrm{diag}(2/3,-1/3,-1/3)$
is the quark charge matrix in flavor space.
$\lambda^a$ are the Gell-Mann matrices in flavor space, with $\lambda^0=\sqrt{2/3}I$. $G$ and $K$ denote the coupling constants for the scalar-pseudoscalar four-quark interaction
and the $U_A(1)$-breaking 't Hooft six-quark interaction, respectively.

In the presence of a constant magnetic field, the propagator of a constituent quark with mass $M_f$ satisfies
\begin{equation}\label{eq:quark_prop}
\left(i\gamma^{\mu}D_{\mu}-M_f\right)S_f(x,x')=i\delta^{(4)}(x-x').
\end{equation}
In Minkowski space with metric $g_{\mu\nu}=\mathrm{diag}(1,-1,-1,-1)$, the solution of Eq.~\eqref{eq:quark_prop} takes the form
\begin{equation}\label{eq:prop_solution}
S_f(x,x')=e^{i\Phi_f(\bm{r}_{\perp},\bm{r}_{\perp}')}\widetilde{S}_f(x-x'),
\end{equation}
where the index $f$ labels the quark flavor, $\bm{r}_{\perp}=(x,y)$, and
\begin{equation}
\Phi_f(\bm{r}_{\perp},\bm{r}_{\perp}')=\frac{1}{2}Q_f eB(x+x')(y-y')
\end{equation}
is the Schwinger phase~\cite{PhysRev.82.664}. The translation-invariant part of the propagator, $\widetilde{S}_f(x-x')$, in the proper-time representation reads
\begin{equation}\label{eq:prop_ft}
\widetilde{S}_f(x-x')=\int\frac{d^4p}{(2\pi)^4}e^{-ip\cdot(x-x')}\widetilde{S}_f(p),
\end{equation}
where
\begin{align}\label{eq:prop_pt}
\widetilde{S}_f(p)
&=\int_{0}^{\infty}ds\exp\left\{is\left[p_0^2-p_3^2-M_f^2+i\epsilon\right]
-is\bm{p}_{\perp}^2\frac{\tan(sQ_f eB)}{sQ_f eB}\right\}\notag\\
&\quad\times\left[\gamma^{\mu}p_{\mu}+M_f+(p^1\gamma^2-p^2\gamma^1)
\tan(sQ_f eB)\right]
\left[1-\gamma^1\gamma^2\tan(sQ_f eB)\right].
\end{align}
Here, $\bm{p}_{\perp}^2=p_1^2+p_2^2$.

In the mean field approximation, by making use of Eq.~(\ref{eq:prop_solution}) one can obtain the thermodynamic potential density as
\begin{eqnarray}\label{Omega}
  \Omega &=& 2 G\left(\sigma_u^2+\sigma_d^2+\sigma_s^2\right)-4 K \sigma_u \sigma_d \sigma_s \nm \\
  &&+ \frac{N_c}{8 \pi^2} \sum_{f=u,d,s} \int_{0}^{\infty} \frac{d s}{s^2}  \exp\left(-sM_f^2\right)
 \theta_4\left(0, \mathrm{e}^{-\frac{1}{4 sT^2}}\right)\left[\frac{Q_f eB}{\tanh(Q_f eB s)}\right],
\end{eqnarray}
where $M_f$ is the constituent mass for the quark flavor $f$ with
\begin{equation}\label{Eq-M}
\begin{aligned}
M_u &=m_u-4 G \sigma_u+2K \sigma_d \sigma_s, \\
M_d &=m_d-4 G \sigma_d+2K  \sigma_s \sigma_u, \\
M_s &=m_s-4 G \sigma_s+2K  \sigma_u \sigma_d.
\end{aligned}
\end{equation}
And $\sigma_f=\langle \overline{\psi}_f \psi_f \rangle$ is the corresponding quark condensate at finite $B$
\begin{equation}\label{Eq-psi}
 \left\langle\overline{\psi}_f\psi_f\right\rangle=-\mathrm{Tr}\,S_f(x,x)=-2M_f I_1(M_f^2)
\end{equation}
with
\begin{equation}\label{Eq-I1}
I_1(M_f^2)=\frac{N_c}{8 \pi^2} \int_{0}^{\infty} \frac{d s}{s}  \exp\left(-sM_f^2\right)
 \theta_4\left(0, \mathrm{e}^{-\frac{1}{4 sT^2}}\right)\left[\frac{Q_f eB}{\tanh({Q_f eB s})}\right].
\end{equation}
Note that $\theta_4\left(u, q\right)$ is the Jacobi theta function~\cite{book:63138}, and we have
\begin{equation}\label{Eq-theta}
  \theta_4\left(0, \mathrm{e}^{-\frac{1}{4 sT^2}}\right)=\theta_3\left(\frac{\pi}{2}, \mathrm{e}^{-\frac{1}{4 sT^2}}\right)=2\sqrt{\pi s}T\theta_{2}\left(0,e^{-4\pi^2T^2s}\right).
\end{equation}

The mean-field variables ($X=\sigma_u$, $\sigma_d$ and $\sigma_s$) are determined by the stationary conditions (the gap equations)
\begin{equation}\label{gap-EQ}
\frac{\partial \Omega}{\partial X} = 0.
\end{equation}
Following the prescription in Ref.~\cite{Sheng:2021evj}, we employ the proper-time regularization scheme with an ultraviolet cutoff $\Lambda_{\mathrm{UV}}$ and an infrared cutoff $\Lambda_{\mathrm{IR}}$~\cite{Hellstern:1997nv,Ebert:1996vx,Bentz:2001vc}. And we define
the light-quark condensate $\langle\bar{\psi}_l\psi_l\rangle=\langle\bar{u}u\rangle+\langle\bar{d}d\rangle$.

\subsection{\label{sec:three}Meson screening masses and meson susceptibilities}

As discussed in Ref.~\cite{Ding:2026ewc}, $\pi^0$ and $\delta^0$, as well as $\eta$ and $\sigma$, remain $U_A(1)$ partners at nonzero
magnetic field. Hence, in this work, the screening mass difference between the $\pi^0$ and $\delta^0$ mesons is used as an effective order parameter to investigate the breaking and restoration of the axial $U_A(1)$ symmetry in a magnetic field. Note that $\eta$ and $\sigma$ denote the mesons with no strange-quark (ns) component, $\eta_{ns}$ and $\sigma_{ns}$. Moreover, $\pi^0$ and $\sigma$, as well as $\delta^0$ and $\eta$, are chiral partners in the presence of a magnetic field.

We first evaluate the screening masses for the meson channels
$\xi=\pi^0,\delta^0,\eta,\sigma$, following Ref.~\cite{Ishii:2013kaa}.
The interpolating current for the meson channel $\xi$ is
\begin{equation}
J_\xi(x)=\bar{\psi}(x)\Gamma_\xi\psi(x)
-\langle\bar{\psi}(x)\Gamma_\xi\psi(x)\rangle,
\end{equation}
where
$\Gamma_{\pi^0}=i\gamma_5\lambda_3$,
$\Gamma_{\delta^0}=\lambda_3$,
$\Gamma_{\eta}=i\gamma_5\left(\sqrt{\frac{2}{3}}\lambda_0+\sqrt{\frac{1}{3}}\lambda_8\right)$,
and
$\Gamma_{\sigma}=\sqrt{\frac{2}{3}}\lambda_0+\sqrt{\frac{1}{3}}\lambda_8$.

The mesonic correlation function is defined by
\begin{equation}\label{eq:corr_eta}
\eta_{\xi\xi}(x)\equiv
\langle 0|T[J_\xi(x)J_\xi^\dagger(0)]|0\rangle,
\end{equation}
where $T$ denotes the time-ordered product. Its Fourier transform reads
\begin{equation}\label{eq:corr_chi}
\chi_{\xi\xi}(k)=i\int d^4x\,e^{ik\cdot x}
\langle 0|T[J_\xi(x)J_\xi^\dagger(0)]|0\rangle.
\end{equation}
The momentum-space correlation function $\chi_{\xi\xi}(k)$ can be obtained using the random phase approximation (RPA)~\cite{Klevansky:1992qe,Florkowski:1997pi}, and is given by
\begin{equation}\label{eq:RPA}
\chi_{\xi\xi}(k)=\frac{\Pi_\xi(k)}{1-2G_\xi\Pi_\xi(k)},
\end{equation}
where the quark-loop polarization function is
\begin{equation}\label{eq:Pi}
\Pi_\xi(k)\equiv -i\int\frac{d^4p}{(2\pi)^4}
\mathrm{Tr}\left[
\Gamma_\xi\widetilde{S}(p)
\Gamma_\xi\widetilde{S}(p-k)
\right],
\end{equation}
and the effective couplings $G_\xi$ are
\begin{align}
G_{\delta^0}=G_\eta &= G+\frac{1}{2}K\sigma_s,\\
G_{\pi^0}=G_\sigma &= G-\frac{1}{2}K\sigma_s.
\end{align}
Note that the Schwinger phases of the quark and antiquark cancel with each other in the neutral-meson polarization functions.

After performing the Matsubara frequency summation, we obtain the explicit expressions for the polarization functions at finite temperature (for $k_0=0$):
\begin{equation}\label{eq:Pi_explicit}
\Pi_\xi(0,\bm{k}_{\perp}^2,k_3^2)
=\sum_{f=u,d}\left[
2I_1(M_f^2)
+\left(k_3^2+4M_f^2\epsilon_\xi\right)I_{2,\parallel}(0,\bm{k}_{\perp}^2,k_3^2,M_f^2)
+\bm{k}_{\perp}^2 I_{2,\perp}(0,\bm{k}_{\perp}^2,k_3^2,M_f^2)
\right],
\end{equation}
where $\epsilon_\xi$ is defined by
\begin{equation}\label{eq:epsilon_xi}
\epsilon_\xi\equiv
\begin{cases}
1, & \xi=\delta^0,\sigma,\\
0, & \xi=\pi^0,\eta,
\end{cases}
\end{equation}
and
\begin{widetext}
\begin{align}\label{eq:I2_parallel}
&I_{2,\parallel}(0,\bm{k}_{\perp}^2,k_3^2,M_f^2)\notag\\
&=-\frac{N_c}{8\pi^2}\int_0^\infty ds\int_0^1 du\,
\exp\left[-sM_f^2-\frac{s(1-u^2)}{4}k_3^2
-\frac{\cosh(sQ_f eB)-\cosh(suQ_f eB)}
{2Q_f eB\sinh(sQ_f eB)}\bm{k}_{\perp}^2\right]\notag\\
&\quad\times\theta_4\!\left(0,e^{-\frac{1}{4sT^2}}\right)
\left[\frac{Q_f eB}{\tanh(sQ_f eB)}\right],
\end{align}
and
\begin{align}\label{eq:I2_perp}
&I_{2,\perp}(0,\bm{k}_{\perp}^2,k_3^2,M_f^2)\notag\\
&=-\frac{N_c}{8\pi^2}\int_0^\infty ds \int_0^1 du\,
\exp\left[-sM_f^2-\frac{s(1-u^2)}{4}k_3^2
-\frac{\cosh(sQ_f eB)-\cosh(suQ_f eB)}
{2Q_f eB\sinh(sQ_f eB)}\bm{k}_{\perp}^2\right]\notag\\
&\quad\times\theta_4\!\left(0,e^{-\frac{1}{4sT^2}}\right)
\left[\frac{Q_f eB\cosh(suQ_f eB)}{\sinh(sQ_f eB)}\right].
\end{align}
\end{widetext}

\begin{figure}
%\begin{tabular}{ccccc}
 \centerline{\includegraphics[scale=0.22]{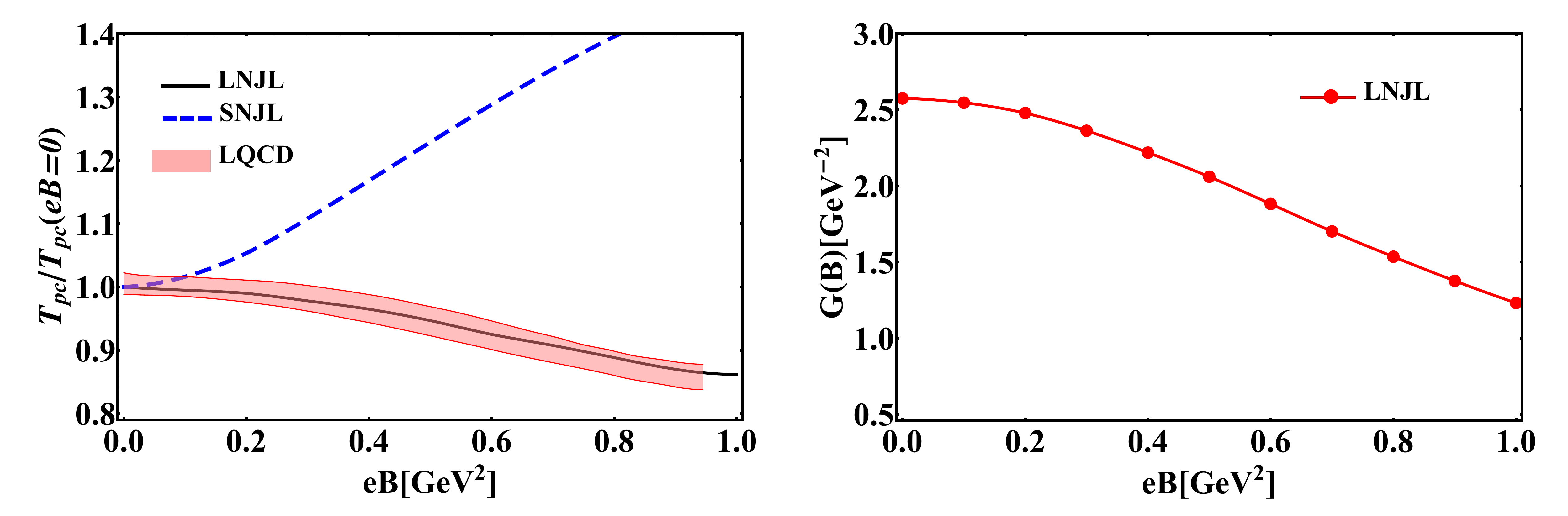}}
%\end{tabular}
\caption{(color online) Left: Normalized pseudo-critical temperature for chiral symmetry as a function of $eB$ in the SNJL model (dashed lines), the LNJL model (solid lines), and the LQCD simulations of Ref.~\cite{Bali:2011qj}. The LQCD data are normalized by $T_{\mathrm{pc}}(eB=0)\mid_{\mathrm{LQCD}}=158~\mathrm{MeV}$, and the NJL results by $T_{\mathrm{pc}}(eB=0)\mid_{\mathrm{NJL}}=156.4~\mathrm{MeV}$. Right: The magnetic-field-dependent coupling constant as a function of $eB$, inferred from the solid curves in the left panel.}
\label{fig1}
\end{figure}

\begin{figure}
%\begin{tabular}{ccccc}
 \centerline{\includegraphics[scale=0.22]{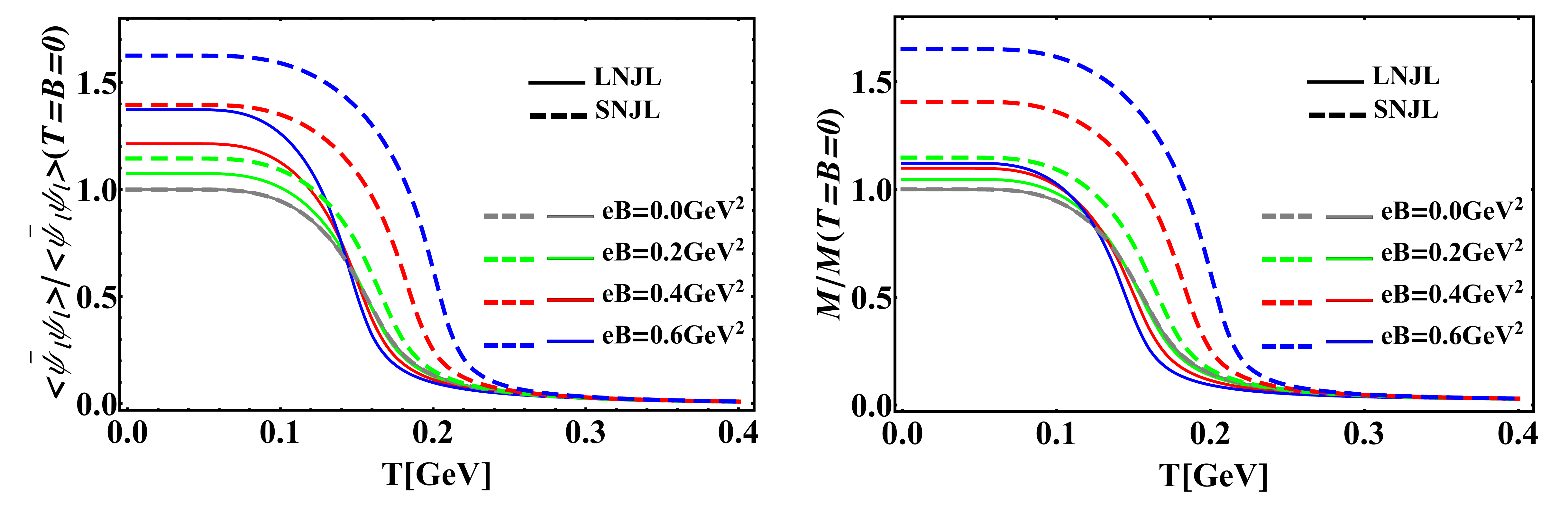}}
%\end{tabular}
\caption{(color online) Left: Normalized light-quark condensate as a function of $T$ at $eB=0.0$, $0.2$, $0.4$ and $0.6\ \mathrm{GeV}^2$ in the SNJL model (dashed lines) and the LNJL model (solid lines). Right: Normalized average constituent quark mass as a function of $T$ at $eB=0.0$, $0.2$, $0.4$ and $0.6\ \mathrm{GeV}^2$ in the SNJL model (dashed lines) and the LNJL model (solid lines).}
\label{fig2}
\end{figure}

At finite temperature and magnetic field, the longitudinal and transverse screening masses of the mesons, $m_{\xi,\mathrm{scr},\parallel}$ and $m_{\xi,\mathrm{scr},\perp}$, are determined by
\begin{equation}\label{eq:m_parallel}
\left[1-2G_\xi \Pi_\xi(0,0,k_3^2)\right]\big|_{k_3^2=-m_{\xi,\mathrm{scr},\parallel}^2}=0,
\end{equation}
and
\begin{equation}\label{eq:m_perp}
\left[1-2G_\xi \Pi_\xi(0,\bm{k}_{\perp}^2,0)\right]\big|_{\bm{k}_{\perp}^2=-m_{\xi,\mathrm{scr},\perp}^2}=0.
\end{equation}
From the expressions for $I_{2,\parallel}$ and $I_{2,\perp}$, it is easy to find that the longitudinal and transverse screening masses generally differ at nonvanishing magnetic field. At $eB=0$, they coincide as a consequence of
spatial rotational symmetry.

As in the LQCD simulations of Ref.~\cite{Ding:2026ewc}, the meson susceptibilities $\chi_{\xi}$ for $\xi=\pi^0$, $\delta^0$, $\eta$ and $\sigma$ are related to the Matsubara Green's function
$\chi^E_{{\xi\xi}} (q_4^2,\bm{q}^2)$ in the momentum representation as
\begin{equation}
\chi_{\xi} =\chi_{\xi \xi}^{\mathrm{E}} (q_4^2, \bm{q}^2) \big|_{q_4 = 0, \bm{q} = 0}.
\end{equation}
Note that in Ref.~\cite{Buchoff:2013nra}, an extra factor $1/2$ is introduced to define the $\chi_{\xi}$ as
single-flavor quantities, and it is not included in our definition.

\section{Numerical results}

\subsection{\label{sec:four,one}Model parameters and gap equations}

\begin{figure}
%\begin{tabular}{ccccc}
 \centerline{\includegraphics[scale=0.20]{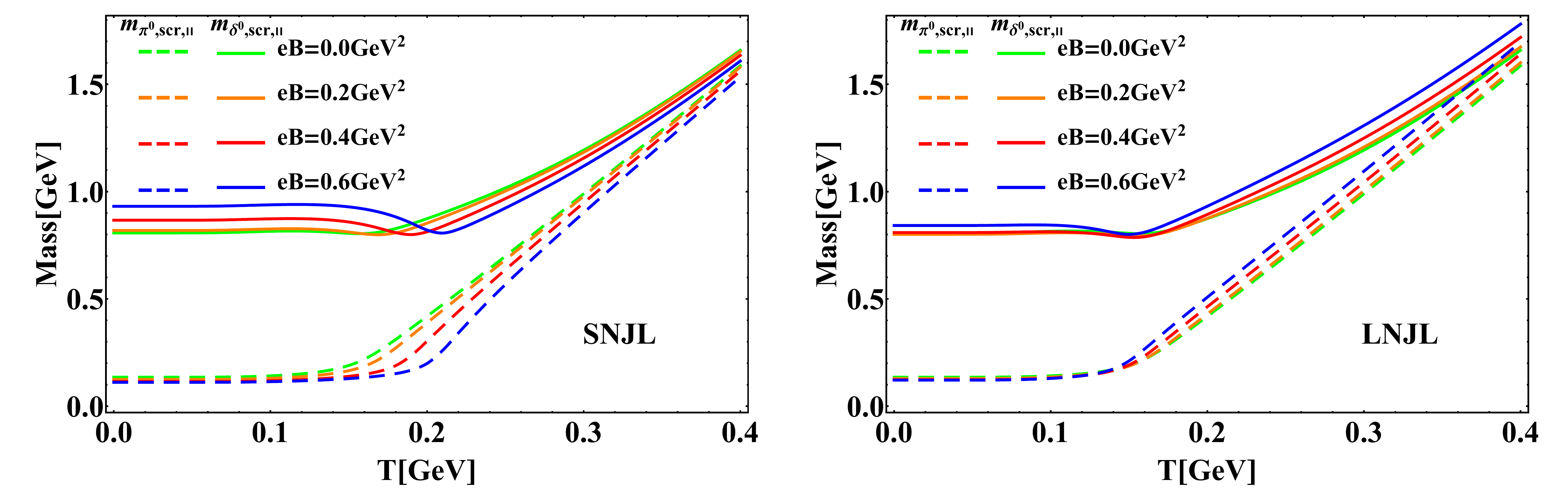}}
 \centerline{(a) }
 \smallskip
 \centerline{\includegraphics[scale=0.20]{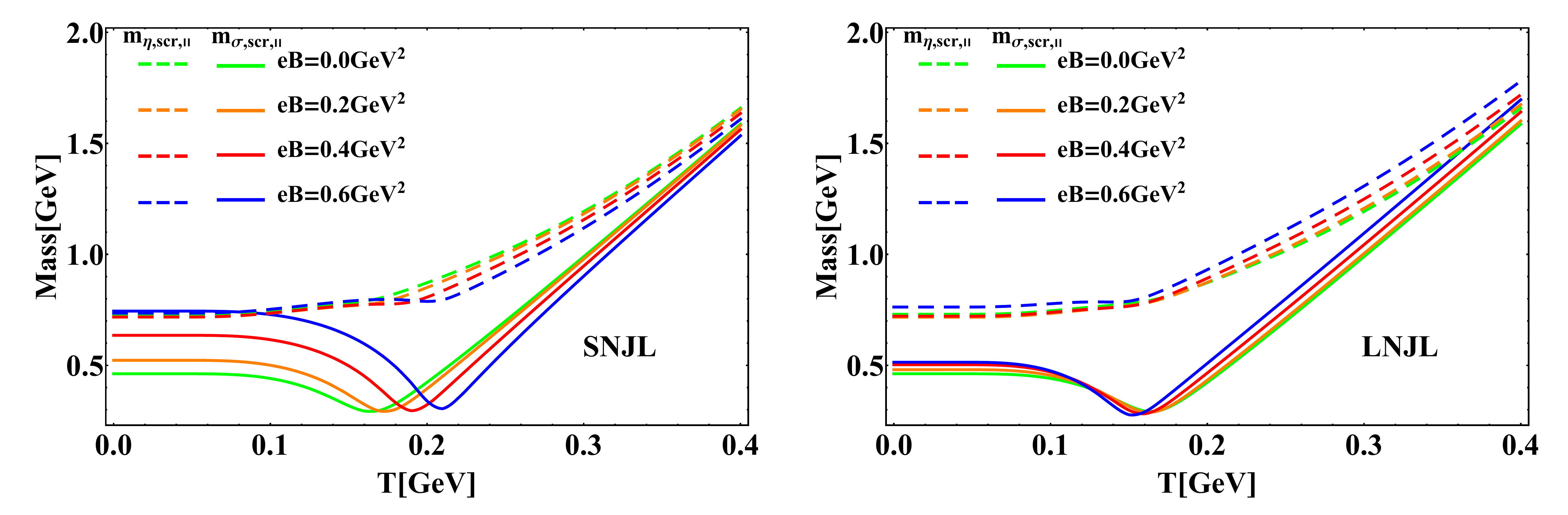}}
 \centerline{(b) }

%\end{tabular}
\caption{(color online) Top: $T$ dependence of the screening masses $m_{\pi^0,\mathrm{scr},\parallel}$ (dashed lines) and $m_{\delta^0,\mathrm{scr},\parallel}$ (solid lines) at $eB=0.0$, $0.2$, $0.4$ and $0.6\ \mathrm{GeV}^2$ in the SNJL model (left) and the LNJL model (right). Bottom: $T$ dependence of the screening masses $m_{\eta,\mathrm{scr},\parallel}$ (dashed lines) and $m_{\sigma,\mathrm{scr},\parallel}$ (solid lines) at $eB=0.0$, $0.2$, $0.4$ and $0.6\ \mathrm{GeV}^2$ in the SNJL model (left) and the LNJL model (right).}
\label{fig3}
\end{figure}

For our numerical calculations, we choose the proper-time regularization scheme, and employ the parameter set $m_u=m_d=5.0$ MeV, $m_s=155.8$ MeV, the ultraviolet cutoff $\Lambda_{{UV}}=1.05$ GeV, $G\Lambda_{{UV}}^2=2.860$, $K\Lambda_{{UV}}^5=105.2$, which are determined by fixing the conditions $m_{{\pi}}=135$ MeV, $f_{{\pi}}=93$ MeV, $m_K=497.7$ MeV and $m_{\eta'}=957.8$ MeV, while $m_u$ and $m_d$ are fixed at 5.0 MeV. Besides, an infrared cutoff $\Lambda_{\mathrm{IR}}=240\ \mathrm{MeV}$ is introduced to mimic confinement and eliminate unphysical quark-antiquark thresholds.
This parameter set yields constituent quark masses
$M_u=M_d=222.5\ \mathrm{MeV}$ and
$M_s=425.6\ \mathrm{MeV}$ in vacuum.

As discussed in Sec.~I, a magnetic-field-dependent coupling
$G(B)$ provides a phenomenological way to incorporate IMC
near the chiral crossover in the NJL model. Following Ref.~\cite{Ferreira:2014kpa}, we determine $G(B)$ by fitting
the magnetic-field dependence of the normalized chiral
pseudo-critical temperature obtained in
LQCD simulations~\cite{Bali:2011qj}, as shown in the left panel of Fig.~\ref{fig1}. The fitted coupling $G(B)$ decreases with increasing magnetic field strength, as shown in the right panel of Fig.~\ref{fig1}, in qualitative agreement with the results of Ref.~\cite{Ferreira:2014kpa}.

In Fig.~\ref{fig2}, we show the normalized light-quark condensate and the normalized average constituent quark mass as functions of temperature at different values of $eB$. Using the inflection points of the quark condensate curves, we define the chiral pseudo-critical temperature $T_{\mathrm{pc}}$. In the SNJL model, the quark condensate is enhanced by the magnetic field (i.e., the MC effect) at all temperatures, and the chiral
pseudo-critical temperature $T_{\mathrm{pc}}$ increases
with $eB$. In contrast, in the LNJL model, the quark condensate exhibits MC at low temperatures and IMC at high temperatures, and $T_{\mathrm{pc}}$ decreases with $eB$, consistent with the LQCD results of Refs.~\cite{Bali:2011qj,Bali:2012zg}. The average constituent quark mass exhibits a similar qualitative magnetic-field dependence. Note that its dependence on $eB$ is different from that in the LNJL model with $G(B)$ inferred from lattice baryon masses~\cite{Endrodi:2019whh}, where $M(B)$ decreases with increasing magnetic field at arbitrary temperatures.

\subsection{Meson screening masses}

\begin{figure}
%\begin{tabular}{ccccc}
 \centerline{\includegraphics[scale=0.20]{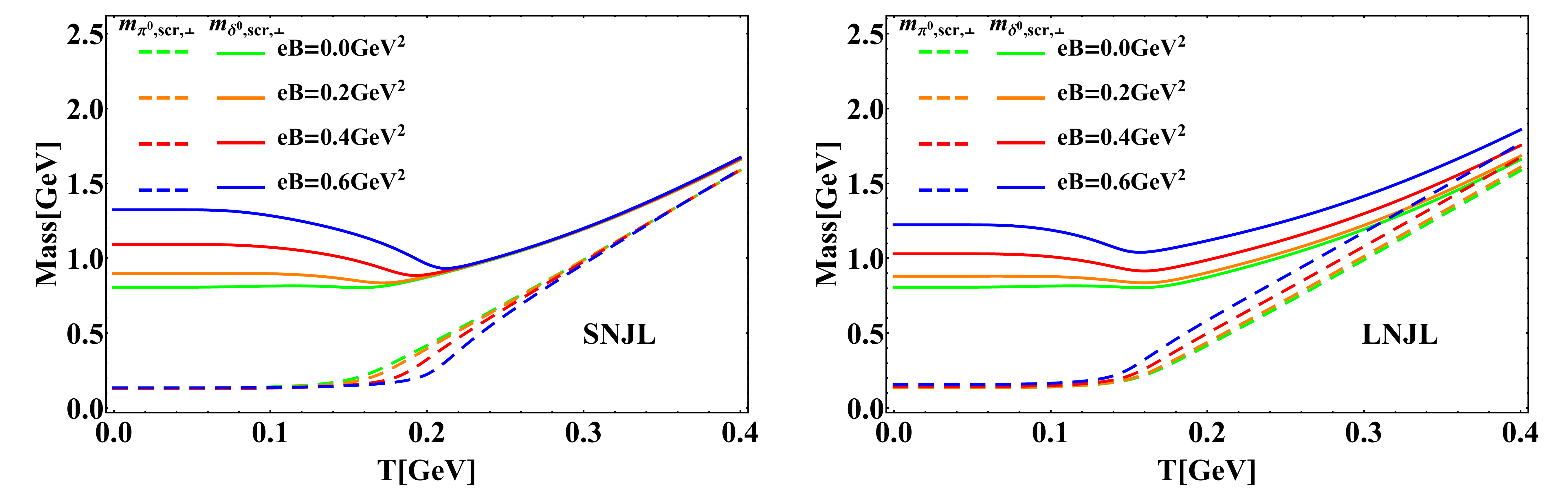}}
 \centerline{(a) }
 \smallskip
 \centerline{\includegraphics[scale=0.20]{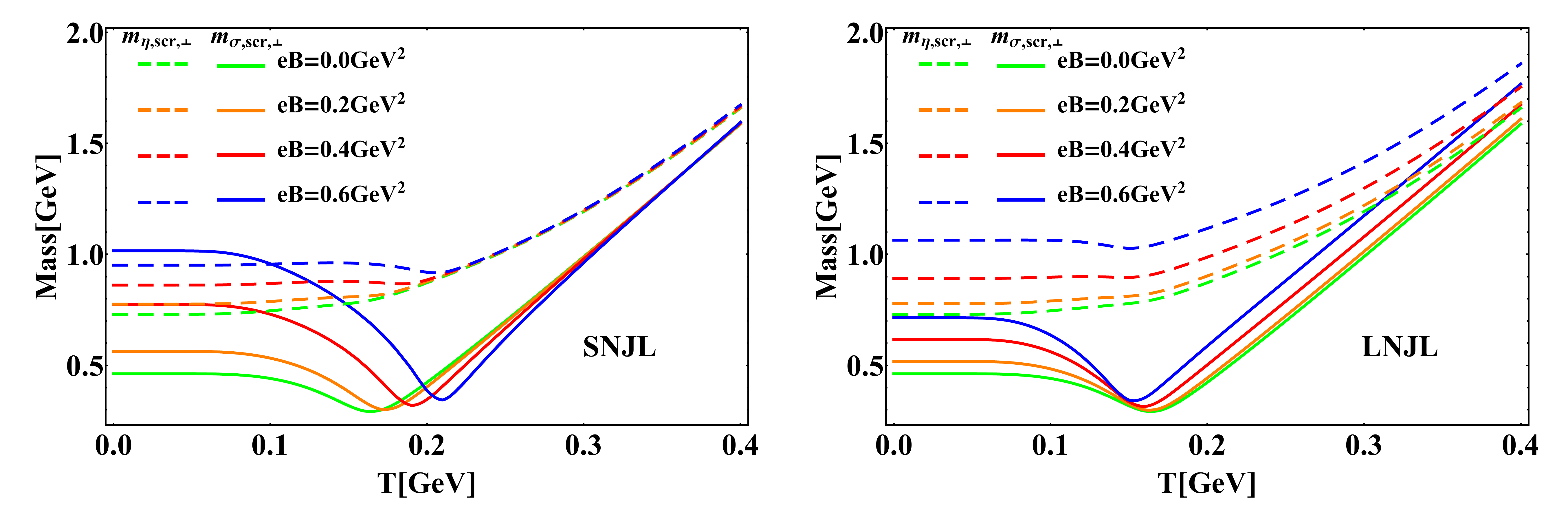}}
 \centerline{(b) }

%\end{tabular}
\caption{(color online) Top: $T$ dependence of the screening masses $m_{\pi^0,\mathrm{scr},\perp}$ (dashed lines) and $m_{\delta^0,\mathrm{scr},\perp}$ (solid lines) at $eB=0.0$, $0.2$, $0.4$ and $0.6\ \mathrm{GeV}^2$ in the SNJL model (left) and the LNJL model (right). Bottom: $T$ dependence of the screening masses $m_{\eta,\mathrm{scr},\perp}$ (dashed lines) and $m_{\sigma,\mathrm{scr},\perp}$ (solid lines) at $eB=0.0$, $0.2$, $0.4$ and $0.6\ \mathrm{GeV}^2$ in the SNJL model (left) and the LNJL model (right).}
\label{fig4}
\end{figure}

Figures~\ref{fig3} and~\ref{fig4} show the temperature dependence of the longitudinal and transverse meson screening masses, respectively, for the $U_A(1)$ partners $\pi^0$ and $\delta^0$, as well as the $U_A(1)$ partners $\eta$ and $\sigma$, at fixed values of $eB$. At low temperatures, the screening masses of all four meson channels in both directions remain approximately constant with increasing temperature. As the temperature approaches $T_{\mathrm{pc}}(B)$, the screening masses of $\pi^0$ and $\eta$ begin to increase rapidly with $T$, whereas those of $\delta^0$ and $\sigma$ first decrease rapidly and then increase as $T$ increases further. For the SNJL model with a constant coupling $G$, larger values of $eB$ shift these rapid changes to higher temperatures. In the LNJL model
with a magnetic-field-dependent coupling $G(B)$, they
instead shift downward. These trends are consistent
with the magnetic-field dependence of
$T_{\mathrm{pc}}(B)$ in the two models. At high temperatures, in both directions, the screening masses of $\pi^0$ and $\delta^0$, as well as those of $\eta$ and $\sigma$, approach each other as the temperature increases. This behavior suggests a tendency toward effective axial $U_A(1)$ symmetry restoration as the temperature increases. Nevertheless, the screening masses of these $U_A(1)$ partners remain nondegenerate even at $T = 400~\mathrm{MeV}$.

\begin{figure}
%\begin{tabular}{ccccc}
 \centerline{\includegraphics[scale=0.20]{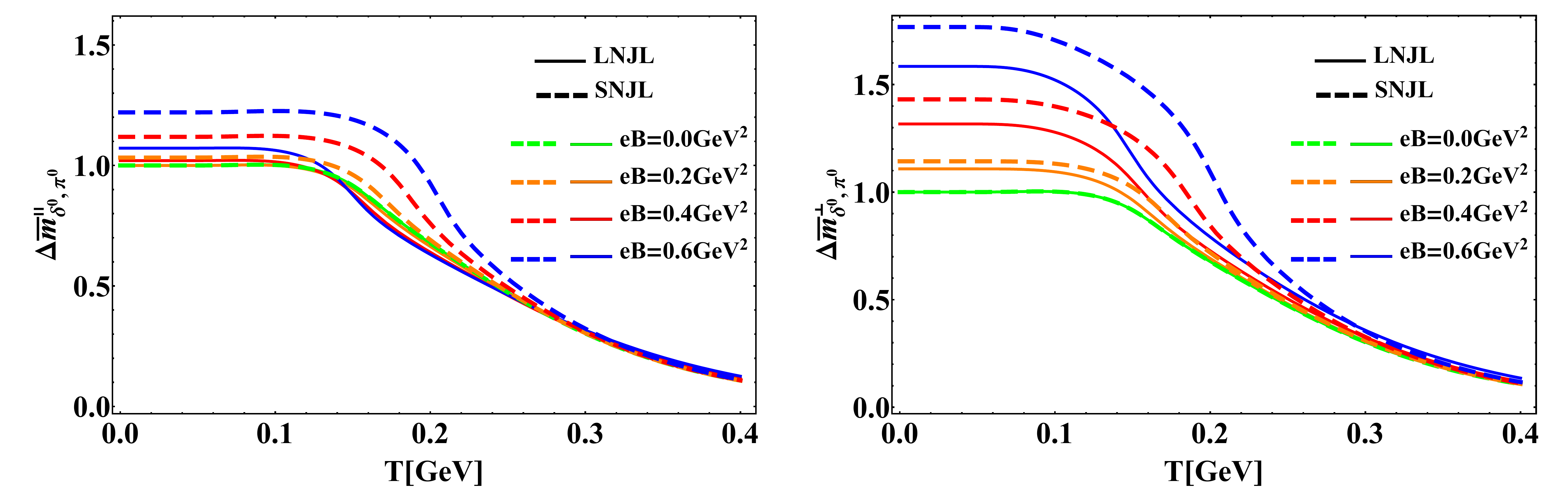}}

%\end{tabular}
\caption{(color online) Left: $T$ dependence of the normalized longitudinal screening mass differences $\Delta\overline{m}_{\delta^0,\pi^0}^{{\parallel}}$ at $eB=0.0$, $0.2$, $0.4$ and $0.6\ \mathrm{GeV}^2$ in the SNJL model (dashed lines) and the LNJL model (solid lines). Right: $T$ dependence of the normalized transverse screening mass differences $\Delta\overline{m}_{\delta^0,\pi^0}^{\perp}$ at $eB=0.0$, $0.2$, $0.4$ and $0.6\ \mathrm{GeV}^2$ in the SNJL model (dashed lines) and the LNJL model (solid lines).}
\label{fig5}
\end{figure}

\begin{figure}
%\begin{tabular}{ccccc}
 \centerline{\includegraphics[scale=0.20]{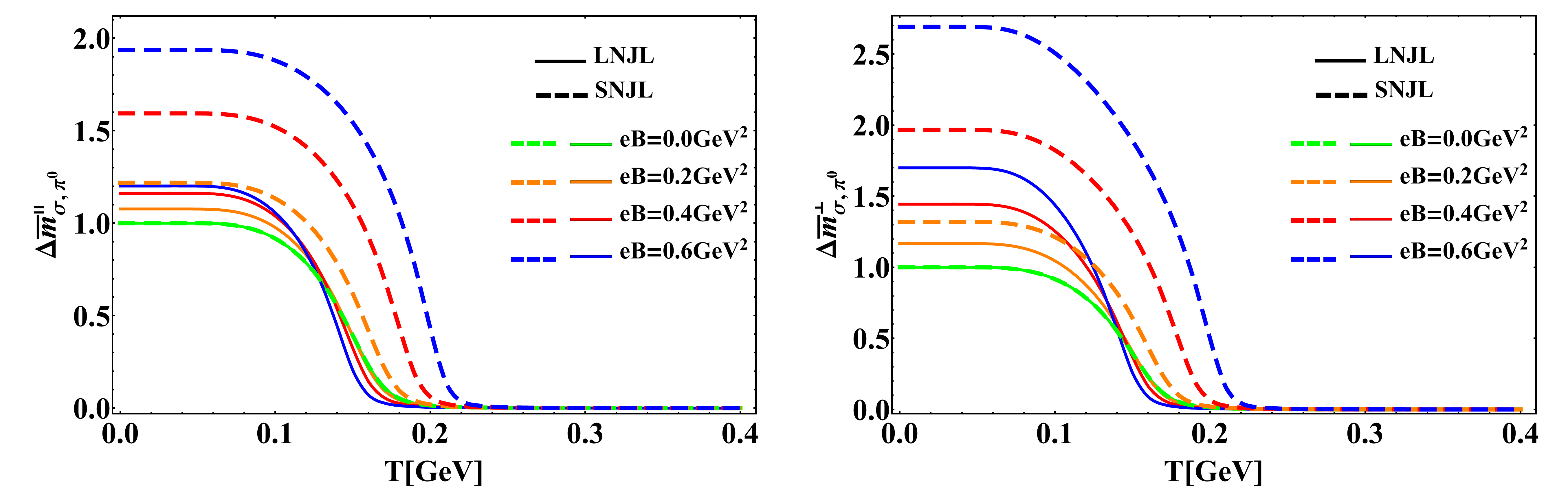}}

%\end{tabular}
\caption{(color online) Left: $T$ dependence of the normalized longitudinal screening mass differences $\Delta\overline{m}_{\sigma,\pi^0}^{{\parallel}}$ at $eB=0.0$, $0.2$, $0.4$ and $0.6\ \mathrm{GeV}^2$ in the SNJL model (dashed lines) and the LNJL model (solid lines). Right: $T$ dependence of the normalized transverse screening mass differences $\Delta\overline{m}_{\sigma,\pi^0}^{\perp}$ at $eB=0.0$, $0.2$, $0.4$ and $0.6\ \mathrm{GeV}^2$ in the SNJL model (dashed lines) and the LNJL model (solid lines).}
\label{fig6}
\end{figure}

\begin{figure}
%\begin{tabular}{ccccc}
 \centerline{\includegraphics[scale=0.20]{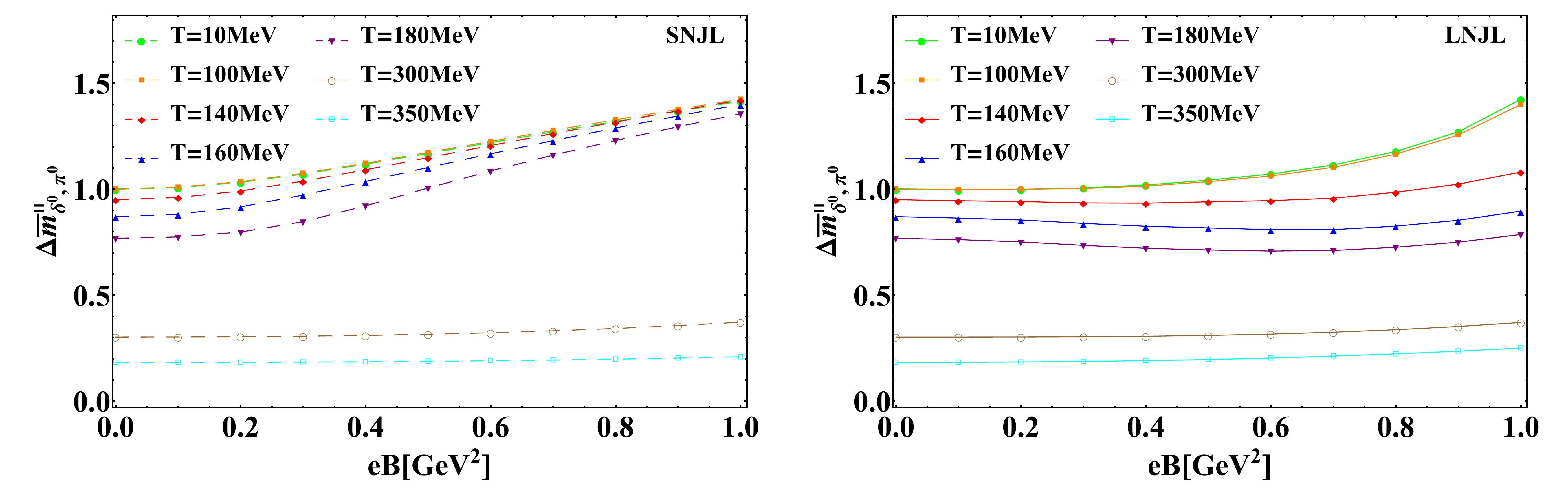}}
 \centerline{(a) }
 \smallskip
 \centerline{\includegraphics[scale=0.20]{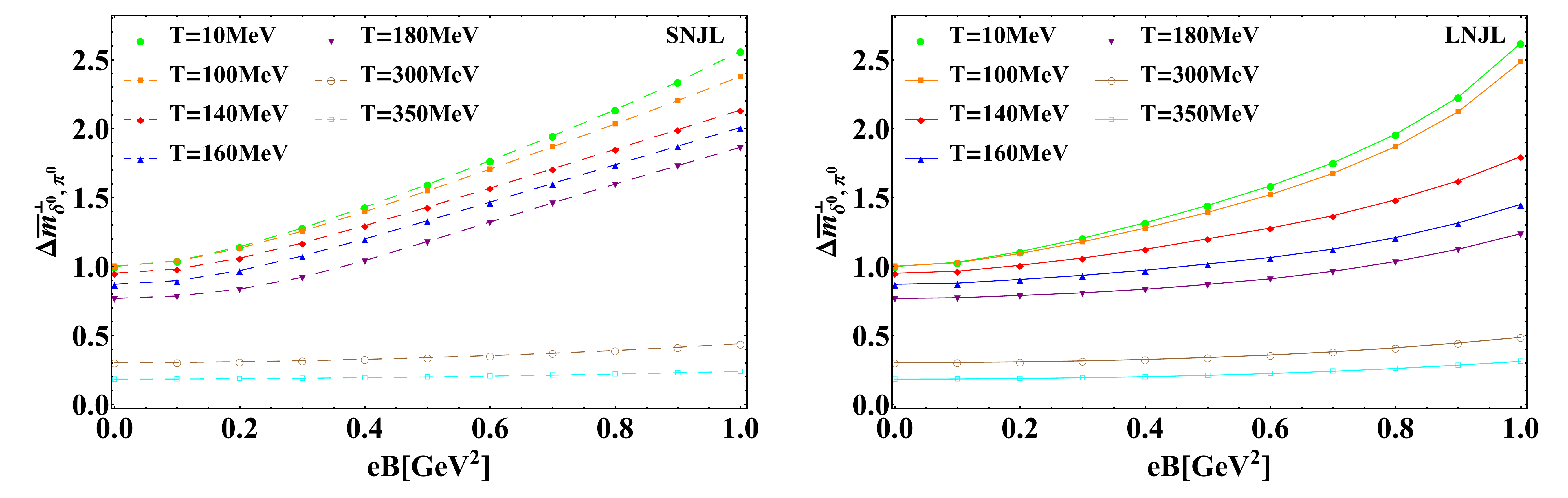}}
 \centerline{(b) }

%\end{tabular}
\caption{(color online) Top: $eB$ dependence of the normalized longitudinal screening mass differences $\Delta\overline{m}_{\delta^0,\pi^0}^{{\parallel}}$ at $T=10$, $100$, $140$, $160$, $180$, $300$ and $350\ \mathrm{MeV}$ in the SNJL model (left) and the LNJL model (right). Bottom: $eB$ dependence of the normalized transverse screening mass differences $\Delta\overline{m}_{\delta^0,\pi^0}^{\perp}$ at $T=10$, $100$, $140$, $160$, $180$, $300$ and $350\ \mathrm{MeV}$ in the SNJL model (left) and the LNJL model (right).}
\label{fig7}
\end{figure}

\begin{figure}
%\begin{tabular}{ccccc}
 \centerline{\includegraphics[scale=0.20]{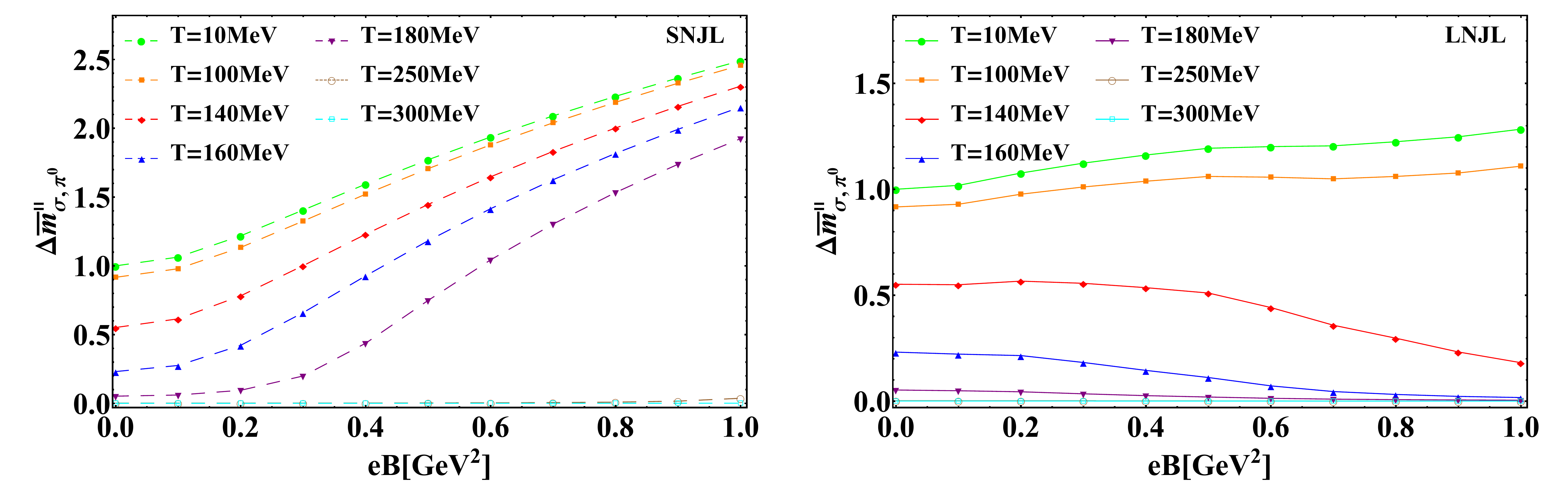}}
 \centerline{(a) }
 \smallskip
 \centerline{\includegraphics[scale=0.20]{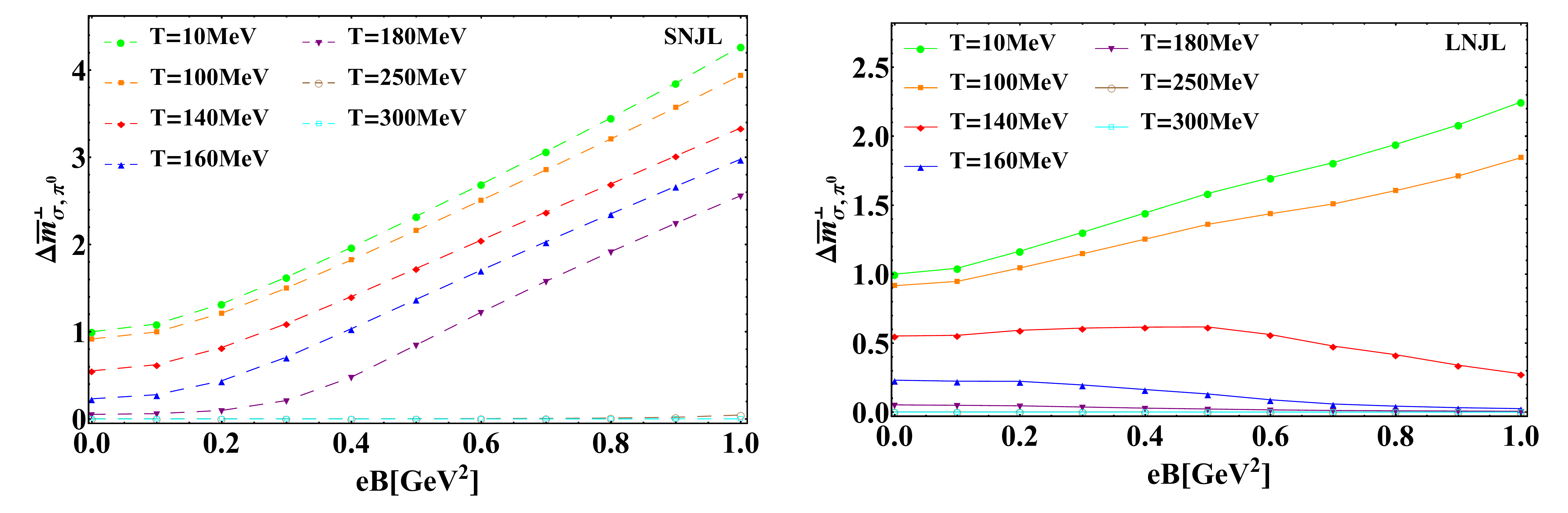}}
 \centerline{(b) }

%\end{tabular}
\caption{(color online)   Top: $eB$ dependence of the normalized longitudinal screening mass differences $\Delta\overline{m}_{\sigma,\pi^0}^{{\parallel}}$ at $T=10$, $100$, $140$, $160$, $180$, $250$ and $300\ \mathrm{MeV}$ in the SNJL model (left) and the LNJL model (right). Bottom: $eB$ dependence of the normalized transverse screening mass differences $\Delta\overline{m}_{\sigma,\pi^0}^{\perp}$ at $T=10$, $100$, $140$, $160$, $180$, $250$ and $300\ \mathrm{MeV}$ in the SNJL model (left) and the LNJL model (right).}
\label{fig8}
\end{figure}

As suggested in Ref.~\cite{Sheng:2021evj}, the screening mass difference between the chiral partners $\pi^0$ and $\sigma$ can serve as an effective order parameter for chiral symmetry, complementing the chiral condensate. By analogy, in this work, we introduce the meson screening mass difference between the $U_A(1)$ partners $\pi^0$ and $\delta^0$ as an alternative order parameter for $U_A(1)$ symmetry. Explicitly, following Ref.~\cite{Sheng:2021evj}, the normalized longitudinal and transverse screening mass differences between $\pi^0$ and $\delta^0$ for $U_A(1)$ symmetry, as well as those between $\pi^0$ and $\sigma$ for chiral symmetry, are defined as
\begin{align}
\Delta\overline{m}_{\delta^0,\pi^0}^{\parallel}(B,T)
&=\frac{m_{\delta^0,\mathrm{scr},\parallel}(B,T)-m_{\pi^0,\mathrm{scr},\parallel}(B,T)}
{m_{\delta^0,\mathrm{scr},\parallel}(0,0)-m_{\pi^0,\mathrm{scr},\parallel}(0,0)},\label{eq:dm_delta_para}\\
\Delta\overline{m}_{\delta^0,\pi^0}^{\perp}(B,T)
&=\frac{m_{\delta^0,\mathrm{scr},\perp}(B,T)-m_{\pi^0,\mathrm{scr},\perp}(B,T)}
{m_{\delta^0,\mathrm{scr},\perp}(0,0)-m_{\pi^0,\mathrm{scr},\perp}(0,0)},\label{eq:dm_delta_perp}
\end{align}
and
\begin{align}
\Delta\overline{m}_{\sigma,\pi^0}^{\parallel}(B,T)
&=\frac{m_{\sigma,\mathrm{scr},\parallel}(B,T)-m_{\pi^0,\mathrm{scr},\parallel}(B,T)}
{m_{\sigma,\mathrm{scr},\parallel}(0,0)-m_{\pi^0,\mathrm{scr},\parallel}(0,0)},\label{eq:dm_sigma_para}\\
\Delta\overline{m}_{\sigma,\pi^0}^{\perp}(B,T)
&=\frac{m_{\sigma,\mathrm{scr},\perp}(B,T)-m_{\pi^0,\mathrm{scr},\perp}(B,T)}
{m_{\sigma,\mathrm{scr},\perp}(0,0)-m_{\pi^0,\mathrm{scr},\perp}(0,0)},\label{eq:dm_sigma_perp}
\end{align}
respectively. As shown in Figs.~\ref{fig3} and~\ref{fig4},
both screening mass differences between $\eta$ and
$\sigma$ exhibit a nonmonotonic temperature dependence,
with peaks near $T_{\mathrm{pc}}$. We therefore focus on the $\delta^0$-$\pi^0$ pair when analyzing the restoration of $U_A(1)$ symmetry.

In Fig.~\ref{fig5}, we show the normalized screening mass differences between the $U_A(1)$ partners $\delta^0$ and $\pi^0$, $\Delta\overline{m}_{\delta^0,\pi^0}^{\parallel}$ and $\Delta\overline{m}_{\delta^0,\pi^0}^{\perp}$, as functions of temperature at $eB=0.0$, $0.2$, $0.4$ and $0.6~\mathrm{GeV}^2$. For $T \lesssim 100\ \mathrm{MeV}$, both are almost constant. When $T \gtrsim 100\ \mathrm{MeV}$, they decrease sharply with increasing temperature, indicating that $U_A(1)$ symmetry is being restored in this temperature range. More importantly, at fixed temperature, $\Delta\overline{m}_{\delta^0,\pi^0}^{\parallel}$ in the LNJL model increases with magnetic field strength for $T \lesssim 120\ \mathrm{MeV}$, whereas it decreases with $eB$ over a certain interval for $140 \lesssim T \lesssim 250\ \mathrm{MeV}$. In the SNJL model, $\Delta\overline{m}_{\delta^0,\pi^0}^{\parallel}$, as expected, increases with $eB$ at all temperatures. As for $\Delta\overline{m}_{\delta^0,\pi^0}^{\perp}$, in both the SNJL and LNJL models, it is enhanced by the magnetic field at any fixed temperature.

In Fig.~\ref{fig6}, we show the normalized screening mass differences between the chiral partners $\sigma$ and $\pi^0$, $\Delta\overline{m}_{\sigma,\pi^0}^{\parallel}$ and $\Delta\overline{m}_{\sigma,\pi^0}^{\perp}$, as functions of temperature at $eB=0.0$, $0.2$, $0.4$ and $0.6~\mathrm{GeV}^2$. By comparing our results with those in Fig.~\ref{fig2},  we find that the curves of $\Delta\overline{m}_{\sigma,\pi^0}^{\parallel}$ and $\Delta\overline{m}_{\sigma,\pi^0}^{\perp}$ behave like those of $M(B)/M(0)$ and $\langle\bar{\psi}_l\psi_l(B)\rangle/\langle\bar{\psi}_l\psi_l(0)\rangle$, respectively, in both the SNJL and LNJL models. In particular, in the LNJL model, both $\Delta\overline{m}_{\sigma,\pi^0}^{\parallel}$ and $\Delta\overline{m}_{\sigma,\pi^0}^{\perp}$ increase with $eB$ at low temperatures, whereas at high temperatures they decrease with $eB$ over a certain range. Moreover, at low temperatures, the longitudinal mass difference increases more slowly with $eB$ than the transverse one, and the normalized constituent quark mass increases more slowly than the normalized light-quark condensate.

Figure~\ref{fig7} shows
$\Delta\overline{m}_{\delta^0,\pi^0}^{\parallel}$
and $\Delta\overline{m}_{\delta^0,\pi^0}^{\perp}$
as functions of $eB$ at several fixed temperatures. In the LNJL model, the increasing behavior of $\Delta\overline{m}_{\delta^0,\pi^0}^{\parallel}$ with $eB$ at low temperatures ($T \leq 100$ MeV), indicating axial magnetic catalysis, turns into a dip-like structure in the crossover region ($T= 140$, $160$, $180$ MeV), implying the onset of axial inverse magnetic catalysis, and then into a monotonically increasing dependence on $eB$ at high temperatures ($T=300$, $350$ MeV). By contrast, $\Delta\overline{m}_{\delta^0,\pi^0}^{\perp}$ in the LNJL model increases with $eB$ at all temperatures. As for $\Delta\overline{m}_{\delta^0,\pi^0}^{\parallel}$ and $\Delta\overline{m}_{\delta^0,\pi^0}^{\perp}$ in the SNJL model, both increase with $eB$ over the entire temperature range.

Figure~\ref{fig8} displays $\Delta\overline{m}_{\sigma,\pi^0}^{\parallel}$ and $\Delta\overline{m}_{\sigma,\pi^0}^{\perp}$ as functions of $eB$ at several fixed values of $T$. In the LNJL model, both $\Delta\overline{m}_{\sigma,\pi^0}^{\parallel}$ and $\Delta\overline{m}_{\sigma,\pi^0}^{\perp}$ generally increase with $eB$ at low temperatures ($T \leq 100$ MeV), corresponding to magnetic catalysis. Around $T_{\mathrm{pc}}$ and above ($T \geq 140$ MeV), both exhibit a decreasing dependence on $eB$ within a certain range of the magnetic field, indicating inverse magnetic catalysis. In contrast to $\Delta\overline{m}_{\delta^0,\pi^0}^{\parallel}$ in Fig.~\ref{fig7}, $\Delta\overline{m}_{\sigma,\pi^0}^{\parallel}$ and $\Delta\overline{m}_{\sigma,\pi^0}^{\perp}$ exhibit a hump-like structure in their $eB$ dependence near the crossover (at $T=140~\mathrm{MeV}$). In the SNJL model, both of them increase with $eB$ at all temperatures, as expected.

Following the prescription in Ref.~\cite{Sheng:2021evj}, we define the axial pseudo-critical temperatures $T_{\mathrm{pc},\parallel}^{A}$ and $T_{\mathrm{pc},\perp}^{A}$ for $U_A(1)$ symmetry from the inflection points of the $\Delta\overline{m}_{\delta^0,\pi^0}^{\parallel}$ and $\Delta\overline{m}_{\delta^0,\pi^0}^{\perp}$ curves,
\begin{equation}\label{eq:Tpc_A}
\left.\frac{\partial^2\Delta\overline{m}_{\delta^0,\pi^0}^{\parallel}}{\partial T^2}\right|_{T=T_{\mathrm{pc},\parallel}^{A}}=0,
\qquad
\left.\frac{\partial^2\Delta\overline{m}_{\delta^0,\pi^0}^{\perp}}{\partial T^2}\right|_{T=T_{\mathrm{pc},\perp}^{A}}=0.
\end{equation}
The chiral pseudo-critical temperatures $T_{\mathrm{pc},\parallel}$ and $T_{\mathrm{pc},\perp}$ are defined analogously using the $\Delta\overline{m}_{\sigma,\pi^0}^{\parallel}$ and $\Delta\overline{m}_{\sigma,\pi^0}^{\perp}$ curves,
\begin{equation}\label{eq:Tpc_chiral}
\left.\frac{\partial^2\Delta\overline{m}_{\sigma,\pi^0}^{\parallel}}{\partial T^2}\right|_{T=T_{\mathrm{pc},\parallel}}=0,
\qquad
\left.\frac{\partial^2\Delta\overline{m}_{\sigma,\pi^0}^{\perp}}{\partial T^2}\right|_{T=T_{\mathrm{pc},\perp}}=0.
\end{equation}

The magnetic-field dependence of the axial and chiral pseudo-critical temperatures is shown in Figs.~\ref{fig11} and~\ref{fig12}, respectively, and is discussed in the next subsection.

\subsection{Meson susceptibilities}

\begin{figure}
%\begin{tabular}{ccccc}
 \centerline{\includegraphics[scale=0.20]{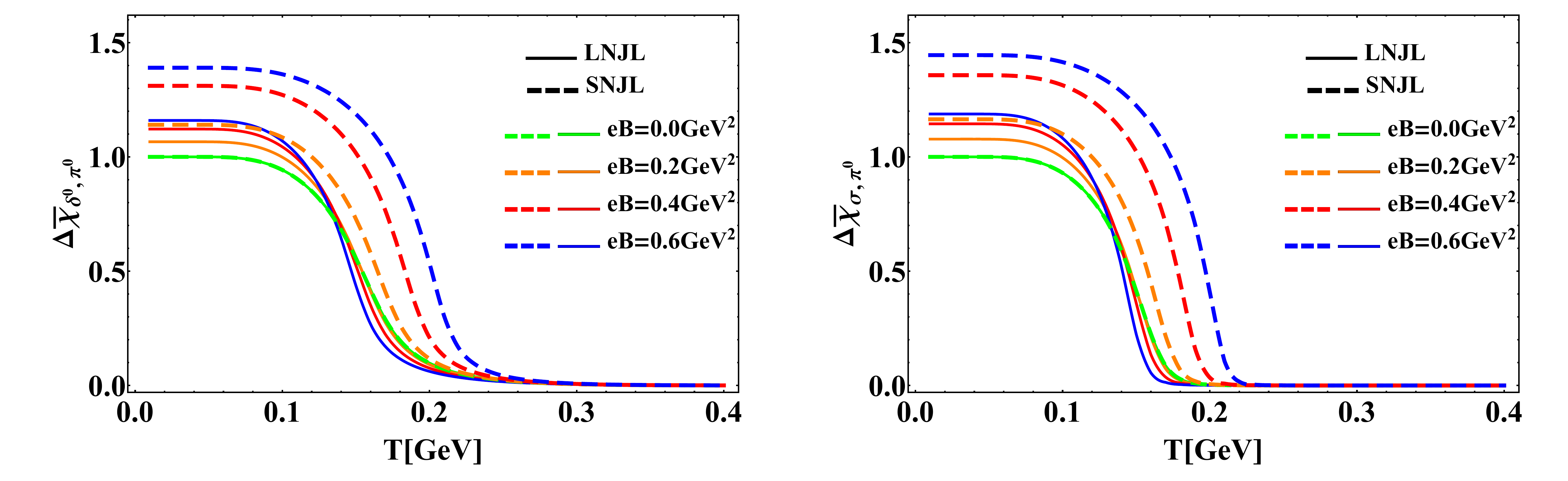}}
%\end{tabular}
\caption{(color online) Left: $T$ dependence of the normalized meson susceptibility differences $\Delta\overline{\chi}_{\delta^0,\pi^0}$ at $eB=0.0$, $0.2$, $0.4$ and $0.6\ \mathrm{GeV}^2$ in the SNJL model (dashed lines) and the LNJL model (solid lines). Right: $T$ dependence of the normalized meson susceptibility differences $\Delta\overline{\chi}_{\sigma,\pi^0}$ at $eB=0.0$, $0.2$, $0.4$ and $0.6\ \mathrm{GeV}^2$ in the SNJL model (dashed lines) and the LNJL model (solid lines).}
\label{fig9}
\end{figure}

\begin{figure}
%\begin{tabular}{ccccc}
 \centerline{\includegraphics[scale=0.20]{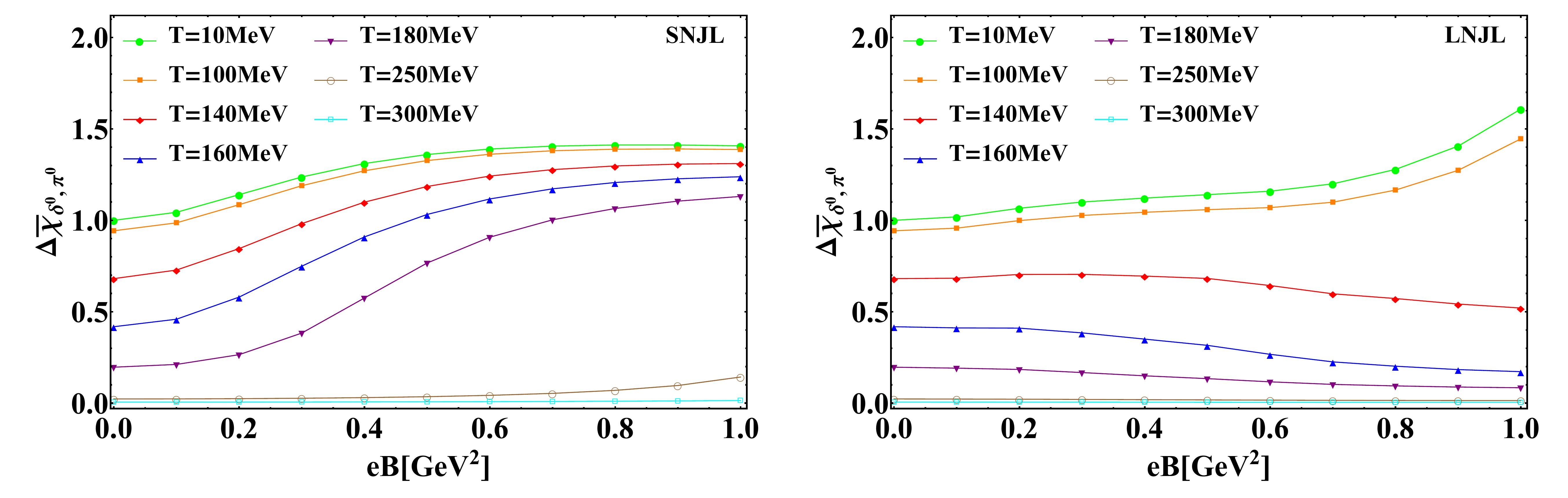}}
 \centerline{(a) }
 \smallskip
 \centerline{\includegraphics[scale=0.20]{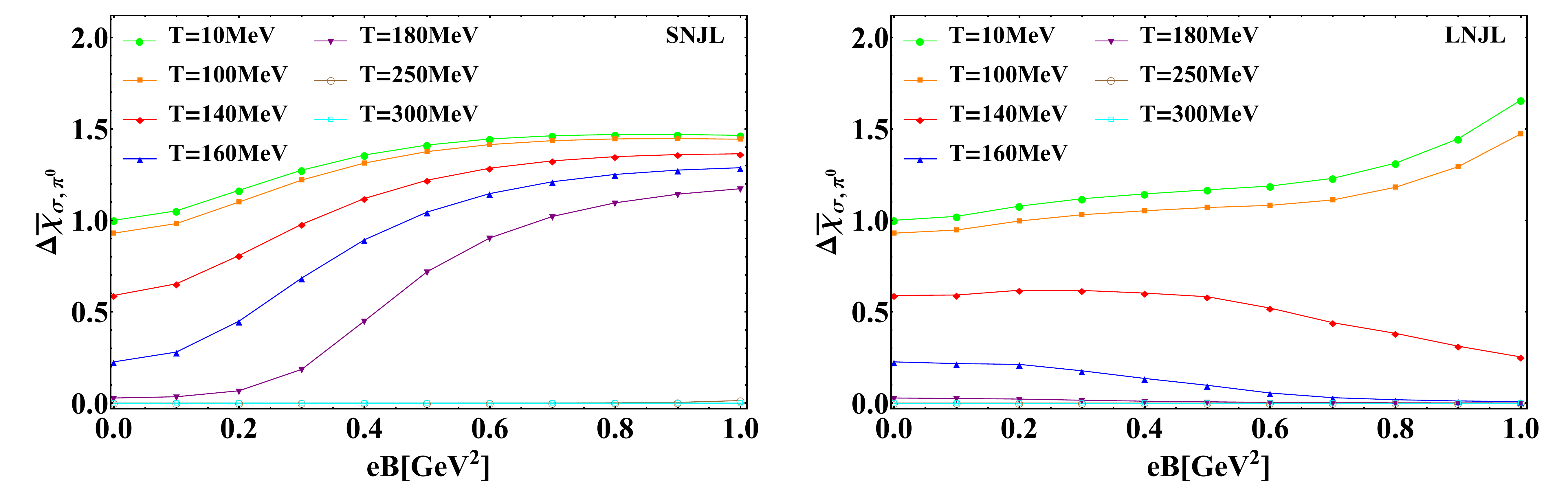}}
 \centerline{(b) }

%\end{tabular}
\caption{(color online)  Top: $eB$ dependence of the normalized meson susceptibility differences $\Delta\overline{\chi}_{\delta^0,\pi^0}$ at $T=10$, $100$, $140$, $160$, $180$, $250$ and $300\ \mathrm{MeV}$ in the SNJL model (left) and the LNJL model (right). Bottom: $eB$ dependence of the normalized meson susceptibility differences $\Delta\overline{\chi}_{\sigma,\pi^0}$ at $T=10$, $100$, $140$, $160$, $180$, $250$ and $300\ \mathrm{MeV}$ in the SNJL model (left) and the LNJL model (right).}
\label{fig10}
\end{figure}

For comparison with the screening mass results,
we define the normalized susceptibility differences
for the $U_A(1)$ and chiral partners as
\begin{equation}\label{eq:chi_delta_pi}
\Delta\overline{\chi}_{\delta^0,\pi^0}(B,T)
=
\frac{
\chi_{\delta^0}(B,T)-\chi_{\pi^0}(B,T)
}{
\chi_{\delta^0}(0,0)-\chi_{\pi^0}(0,0)
},
\end{equation}
and
\begin{equation}\label{eq:chi_sigma_pi}
\Delta\overline{\chi}_{\sigma,\pi^0}(B,T)
=
\frac{
\chi_{\sigma}(B,T)-\chi_{\pi^0}(B,T)
}{
\chi_{\sigma}(0,0)-\chi_{\pi^0}(0,0)
},
\end{equation}
respectively, with each difference normalized to
its vacuum value.

Figure~\ref{fig9} shows the temperature dependence of $\Delta\overline{\chi}_{\delta^0,\pi^0}$ and $\Delta\overline{\chi}_{\sigma,\pi^0}$ at fixed $eB$, while Fig.~\ref{fig10} shows the $eB$ dependence of both quantities at fixed temperatures. In the SNJL model, both differences are enhanced
by the magnetic field, with several curves
becoming approximately flat at larger $eB$
within the plotted range. In the LNJL model, both differences increase with
$eB$ at low temperatures ($T\leq100~\mathrm{MeV}$),
indicating AMC in the
$\delta^0$--$\pi^0$ channel and MC
in the $\sigma$--$\pi^0$ channel.
Around and above the chiral crossover, both decrease
over a finite interval of magnetic-field strength,
indicating AIMC and IMC, respectively. Although these trends are in qualitative agreement with the LQCD results~\cite{Ding:2026ewc},
the values of $eB$ at which both meson susceptibility differences start to decrease are much smaller
in the LNJL model than in LQCD~\cite{Ding:2026ewc}. For instance, at $T=140~\mathrm{MeV}$,
$\Delta\overline{\chi}_{\delta^0,\pi^0}$ begins
to decrease at $eB\approx0.25~\mathrm{GeV}^2$
in the LNJL model, compared with
$eB\approx0.9~\mathrm{GeV}^2$ in LQCD.
Furthermore, in the LNJL model,
$\Delta\overline{\chi}_{\delta^0,\pi^0}$ becomes
close to zero at
$T\approx250~\mathrm{MeV}$, whereas
$\Delta\overline{\chi}_{\sigma,\pi^0}$ does so
at $T\approx200~\mathrm{MeV}$. This implies that the effective restoration temperature of $U_A(1)$ symmetry is higher than that of chiral symmetry at zero and nonzero magnetic field, and such behavior has been reported in LQCD simulations at $eB=0$~\cite{Bhattacharya:2014ara}.

Similarly, we define the pseudo-critical temperatures $T_{\mathrm{pc},\chi}^{A}$ for $U_A(1)$ symmetry and $T_{\mathrm{pc},\chi}$ for chiral symmetry from the inflection points of the $\Delta\overline{\chi}_{\delta^0,\pi^0}$ and $\Delta\overline{\chi}_{\sigma,\pi^0}$ curves shown in Fig.~\ref{fig9},
\begin{equation}
\left.\frac{\partial^2\Delta\overline{\chi}_{\delta^0,\pi^0}}{\partial T^2}\right|_{T=T_{\mathrm{pc},\chi}^{A}}=0
\quad\text{and}\quad
\left.\frac{\partial^2\Delta\overline{\chi}_{\sigma,\pi^0}}{\partial T^2}\right|_{T=T_{\mathrm{pc},\chi}}=0.
\label{eq:Tpc_chi}
\end{equation}

\begin{figure}
%\begin{tabular}{ccccc}
 \centerline{\includegraphics[scale=0.20]{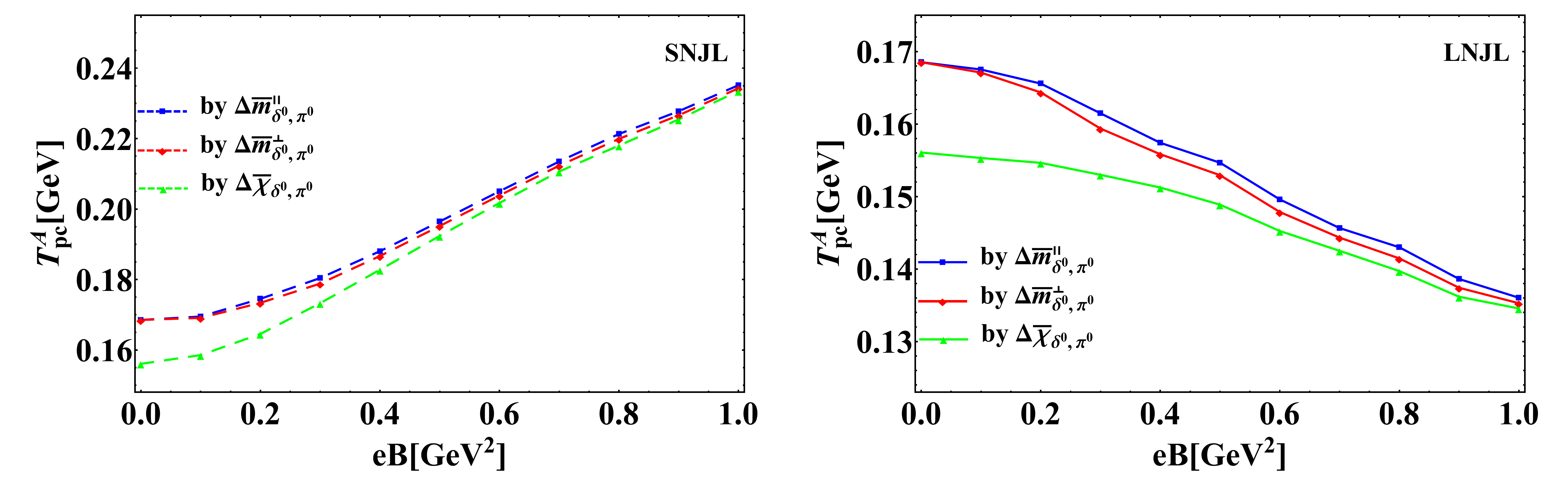}}
%\end{tabular}
\caption{(color online) $eB$ dependence of the pseudo-critical temperatures $T_{\mathrm{pc}}^A$ for $U_A(1)$ symmetry, determined from $\Delta\overline{m}_{\delta^0,\pi^0}^{\parallel}$, $\Delta\overline{m}_{\delta^0,\pi^0}^{\perp}$ and $\Delta\overline{\chi}_{\delta^0,\pi^0}$ in the SNJL model (left) and LNJL model (right).}
\label{fig11}
\end{figure}

\begin{figure}
%\begin{tabular}{ccccc}
 \centerline{\includegraphics[scale=0.20]{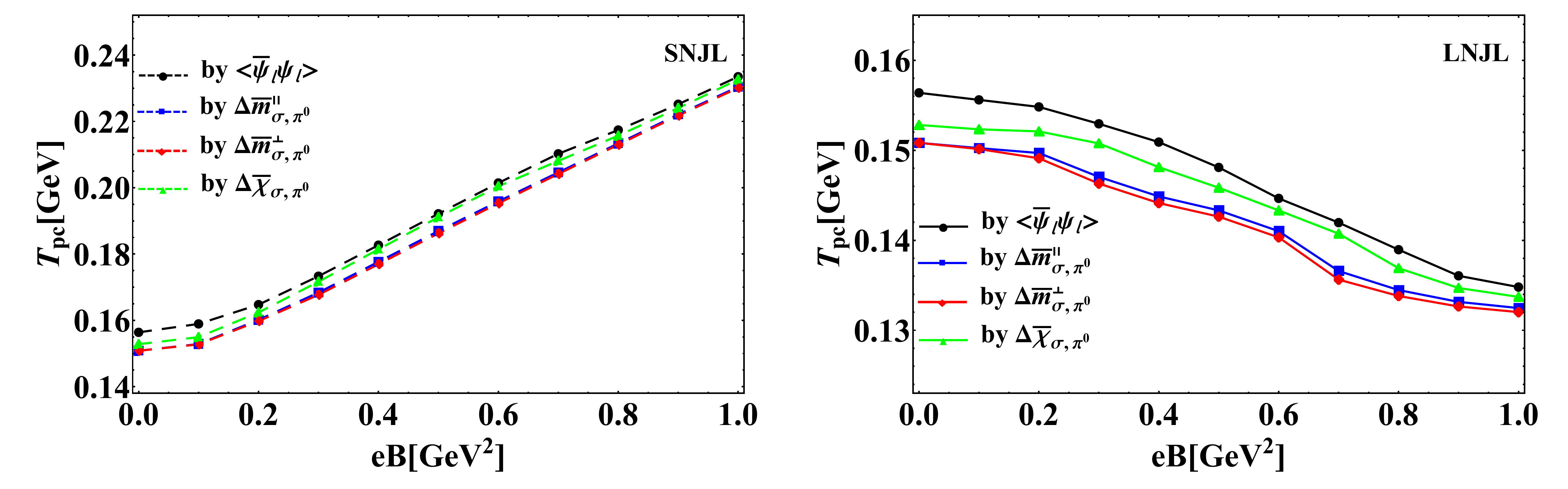}}
%\end{tabular}
\caption{(color online)  $eB$ dependence of the pseudo-critical temperatures $T_{\mathrm{pc}}$ for chiral symmetry, determined from $\langle\bar{\psi_l}\psi_l\rangle$, $\Delta\overline{m}_{\sigma,\pi^0}^{\parallel}$, $\Delta\overline{m}_{\sigma,\pi^0}^{\perp}$ and $\Delta\overline{\chi}_{\sigma,\pi^0}$ in the SNJL model (left) and LNJL model (right).}
\label{fig12}
\end{figure}

Figure~\ref{fig11} shows the $eB$ dependence of the axial pseudo-critical temperatures $T_{\mathrm{pc}}^A$, extracted from $\Delta\overline{m}_{\delta^0,\pi^0}^{\parallel}$, $\Delta\overline{m}_{\delta^0,\pi^0}^{\perp}$ and $\Delta\overline{\chi}_{\delta^0,\pi^0}$. For comparison, Fig.~\ref{fig12} presents the $eB$ dependence of the chiral pseudo-critical temperatures $T_{\mathrm{pc}}$, extracted from $\langle\bar{\psi}_l\psi_l\rangle$, $\Delta\overline{m}_{\sigma,\pi^0}^{\parallel}$, $\Delta\overline{m}_{\sigma,\pi^0}^{\perp}$ and $\Delta\overline{\chi}_{\sigma,\pi^0}$. All extracted pseudo-critical temperatures increase with $eB$ in the SNJL model and decrease in the LNJL model. The latter trend is in qualitative agreement with the $eB$ dependence of $T_{\mathrm{pc}}$ obtained from the quark condensate in LQCD. In particular, in the LNJL model, although $\Delta\overline{m}_{\delta^0,\pi^0}^{\perp}$ increases with $eB$ at all temperatures, $T_{\mathrm{pc},\perp}^A$ is nevertheless suppressed by $eB$. In addition, for each corresponding diagnostic, $T_{\mathrm{pc}}^A$ is higher than $T_{\mathrm{pc}}$ at a given $eB$.

\section{Summary and conclusions}\label{sec:five}

In this work, we have investigated effective axial $U_A(1)$ symmetry restoration in an external magnetic field using meson screening masses and susceptibilities in a $(2+1)$-flavor NJL model. We compared the standard NJL model with a constant
four-quark coupling $G$ and a lattice-improved NJL
model with $G(B)$ fitted to the magnetic-field
dependence of the normalized chiral pseudo-critical
temperature from lattice QCD, while keeping the
't Hooft coupling $K$ fixed. The breaking and restoration of $U_A(1)$ and chiral symmetries are characterized by the normalized longitudinal and transverse screening mass differences, as well as the normalized meson susceptibility differences, for the $\delta^0$-$\pi^0$ and $\sigma$-$\pi^0$ partners, respectively.

The longitudinal and transverse screening mass differences between the $U_A(1)$ partners $\delta^0$ and $\pi^0$ exhibit a directional dependence in their response to $eB$. In the SNJL model, both differences increase with $eB$ at all temperatures. In the LNJL model, the longitudinal mass difference exhibits AMC at low temperatures and develops a nonmonotonic dependence on $eB$ in the crossover region, where its suppression over a certain range of $eB$ signals AIMC. At sufficiently high temperatures, an increasing dependence on $eB$ reappears. The transverse screening mass difference between the $U_A(1)$ partners, however, remains enhanced by $eB$ at all temperatures, even in the LNJL model. This behavior differs from that of the screening mass differences between the chiral partners $\sigma$ and $\pi^0$: in the LNJL model, both the longitudinal and transverse screening mass differences exhibit MC at low temperatures and IMC over a range of $eB$ around and above the chiral crossover.
Additionally, in our model the screening masses of $\delta^0$ and $\pi^0$ remain nondegenerate even at $T=400~\mathrm{MeV}$, both with and without a magnetic field. In contrast, the LQCD simulations of Ref.~\cite{Bazavov:2019www} show that their screening masses become degenerate around $T \approx 200$ MeV at $eB=0$. One possible explanation for this discrepancy is the
absence of an explicit or implicit temperature dependence in the
couplings $G$ and $K$, as suggested by the analysis
in Ref.~\cite{Ishii:2015ira}.
This possibility will be investigated in future work.

The meson susceptibility differences provide complementary probes of these symmetry patterns. In the SNJL model, both the $\delta^0$-$\pi^0$ and $\sigma$-$\pi^0$ meson susceptibility differences increase with $eB$. In the LNJL model, they are enhanced by $eB$ at low temperatures and suppressed over a range of $eB$ around and above the chiral crossover, qualitatively consistent with recent LQCD results~\cite{Ding:2026ewc}.
Near the crossover,
however, the onset of suppression in both the axial
and chiral channels occurs at substantially larger
$eB$ in LQCD than in the LNJL model. Moreover, with increasing temperature, the $\delta^0$-$\pi^0$ meson susceptibility difference nearly vanishes at a higher temperature than the $\sigma$-$\pi^0$ one. This suggests that $U_A(1)$ symmetry breaking tends to survive longer than the chiral symmetry breaking as the temperature grows.

The axial and chiral pseudo-critical temperatures extracted from the thermal inflection points of the screening mass differences increase with $eB$ in the SNJL model, whereas they decrease in the LNJL model. These trends are in good agreement with those obtained from the quark condensate and the meson susceptibility differences. It means that the screening mass differences between the $U_A(1)$ partners and between the chiral partners serve as alternative probes of the breaking and restoration of the corresponding symmetries. In particular, in the LNJL model, we find that the axial pseudo-critical temperature $T_{\mathrm{pc},\perp}^{A}$ extracted from
$\Delta\overline{m}_{\delta^0,\pi^0}^{\perp}$ decreases
with $eB$, even though this screening mass difference increases
with $eB$ at fixed temperature. This illustrates that a decrease in the axial pseudo-critical
temperature does not necessarily imply a suppression of
the transverse screening mass difference $\Delta\overline{m}_{\delta^0,\pi^0}^{\perp}$ by the magnetic
field at fixed temperature.

\begin{acknowledgments}
The authors thank Shinya Matsuzaki for useful discussions.
The work of L. Y. is supported by the
NSFC under Grant No. 11605072 and the Seeds Funding of Jilin University.
The work of X. W. is supported by the National Natural Science
Foundation of China under Grants No. 12675170 and No. 12635010, and by the Anhui University of Science and Technology under Grant No. YJ20240001.
The work of D. L. is supported by the National Natural Science
Foundation of China under Grants No. 12275108 and No. 12235016.
\end{acknowledgments}

\bibliography{bib}% Produces the bibliography via BibTeX.(通过BibTex来产生参考文献)	

@article{Kharzeev:2015kna,
	author = "Kharzeev, Dmitri E.",
	title = "{Topology, magnetic field, and strongly interacting matter}",
	eprint = "1501.01336",
	archivePrefix = "arXiv",
	primaryClass = "hep-ph",
	doi = "10.1146/annurev-nucl-102313-025420",
	journal = "Ann. Rev. Nucl. Part. Sci.",
	volume = "65",
	pages = "193--214",
	year = "2015"
}

@article{Miransky:2015ava,
	author = "Miransky, Vladimir A. and Shovkovy, Igor A.",
	title = "{Quantum field theory in a magnetic field: From quantum chromodynamics to graphene and Dirac semimetals}",
	eprint = "1503.00732",
	archivePrefix = "arXiv",
	primaryClass = "hep-ph",
	doi = "10.1016/j.physrep.2015.02.003",
	journal = "Phys. Rept.",
	volume = "576",
	pages = "1--209",
	year = "2015"
}

@article{Andersen:2014xxa,
	author = "Andersen, Jens O. and Naylor, William R. and Tranberg, Anders",
	title = "{Phase diagram of QCD in a magnetic field: A review}",
	eprint = "1411.7176",
	archivePrefix = "arXiv",
	primaryClass = "hep-ph",
	doi = "10.1103/RevModPhys.88.025001",
	journal = "Rev. Mod. Phys.",
	volume = "88",
	pages = "025001",
	year = "2016"
}

@ARTICLE{Vachaspati:1991nm,
	author = "Vachaspati, T.",
	title = "{Magnetic fields from cosmological phase transitions}",
	doi = "10.1016/0370-2693(91)90051-Q",
	journal = "Phys. Lett. B",
	volume = "265",
	pages = "258--261",
	year = "1991"
}

@ARTICLE{Enqvist:1993np,
	author = "Enqvist, K. and Olesen, P.",
	title = "{On primordial magnetic fields of electroweak origin}",
	eprint = "hep-ph/9308270",
	archivePrefix = "arXiv",
	reportNumber = "NBI-HE-93-33",
	doi = "10.1016/0370-2693(93)90799-N",
	journal = "Phys. Lett. B",
	volume = "319",
	pages = "178--185",
	year = "1993"
}

@article{Duncan:1992hi,
	author = "Duncan, Robert C. and Thompson, Christopher",
	title = "{Formation of very strongly magnetized neutron stars - implications for gamma-ray bursts}",
	doi = "10.1086/186413",
	journal = "Astrophys. J. Lett.",
	volume = "392",
	pages = "L9",
	year = "1992"
}

@article{Skokov:2009qp,
	author = "Skokov, V. and Illarionov, A. Yu. and Toneev, V.",
	title = "{Estimate of the magnetic field strength in heavy-ion collisions}",
	eprint = "0907.1396",
	archivePrefix = "arXiv",
	primaryClass = "nucl-th",
	doi = "10.1142/S0217751X09047570",
	journal = "Int. J. Mod. Phys. A",
	volume = "24",
	pages = "5925--5932",
	year = "2009"
}

@article{Voronyuk:2011jd,
	author = "Voronyuk, V. and Toneev, V. D. and Cassing, W. and Bratkovskaya, E. L. and Konchakovski, V. P. and Voloshin, S. A.",
	title = "{(Electro-)Magnetic field evolution in relativistic heavy-ion collisions}",
	eprint = "1103.4239",
	archivePrefix = "arXiv",
	primaryClass = "nucl-th",
	doi = "10.1103/PhysRevC.83.054911",
	journal = "Phys. Rev. C",
	volume = "83",
	pages = "054911",
	year = "2011"
}

@article{Bzdak:2011yy,
	author = "Bzdak, Adam and Skokov, Vladimir",
	title = "{Event-by-event fluctuations of magnetic and electric fields in heavy ion collisions}",
	eprint = "1111.1949",
	archivePrefix = "arXiv",
	primaryClass = "hep-ph",
	reportNumber = "BNL-96541-2011-JA, RBRC-927",
	doi = "10.1016/j.physletb.2012.02.065",
	journal = "Phys. Lett. B",
	volume = "710",
	pages = "171--174",
	year = "2012"
}

@article{Deng:2012pc,
	author = "Deng, Wei-Tian and Huang, Xu-Guang",
	title = "{Event-by-event generation of electromagnetic fields in heavy-ion collisions}",
	eprint = "1201.5108",
	archivePrefix = "arXiv",
	primaryClass = "nucl-th",
	doi = "10.1103/PhysRevC.85.044907",
	journal = "Phys. Rev. C",
	volume = "85",
	pages = "044907",
	year = "2012"
}

@article{Ahmed:2026qzw,
    author = "Ahmed, Hiwa A. and Li, Danning and Kawaguchi, Mamiya and Huang, Mei",
    title = "{Probing the chiral and U(1) axial symmetry restoration via meson susceptibilities in holographic QCD}",
    eprint = "2603.12911",
    archivePrefix = "arXiv",
    primaryClass = "hep-ph",
    doi = "10.1103/qfmp-b7y3",
    journal = "Phys. Rev. D",
    volume = "114",
    number = "2",
    pages = "026032",
    year = "2026"
}

@article{Buividovich:2008wf,
	author = "Buividovich, P. V. and Chernodub, M. N. and Luschevskaya, E. V. and Polikarpov, M. I.",
	title = "{Numerical study of chiral symmetry breaking in non-Abelian gauge theory with background magnetic field}",
	eprint = "0812.1740",
	archivePrefix = "arXiv",
	primaryClass = "hep-lat",
	reportNumber = "ITEP-LAT-2008-23",
	doi = "10.1016/j.physletb.2009.11.017",
	journal = "Phys. Lett. B",
	volume = "682",
	pages = "484--489",
	year = "2010"
}

@article{Braguta:2010ej,
	author = "Braguta, V. V. and Buividovich, P. V. and Kalaydzhyan, T. and Kuznetsov, S. V. and Polikarpov, M. I.",
	title = "{The Chiral Magnetic Effect and chiral symmetry breaking in SU(3) quenched lattice gauge theory}",
	eprint = "1011.3795",
	archivePrefix = "arXiv",
	primaryClass = "hep-lat",
	reportNumber = "DESY-10-171, ITEP-LAT-2010-13",
	doi = "10.1134/S1063778812030052",
	journal = "Phys. Atom. Nucl.",
	volume = "75",
	pages = "488--492",
	year = "2012"
}

@article{DElia:2010abb,
	author = "D'Elia, Massimo and Mukherjee, Swagato and Sanfilippo, Francesco",
	title = "{QCD Phase Transition in a Strong Magnetic Background}",
	eprint = "1005.5365",
	archivePrefix = "arXiv",
	primaryClass = "hep-lat",
	doi = "10.1103/PhysRevD.82.051501",
	journal = "Phys. Rev. D",
	volume = "82",
	pages = "051501",
	year = "2010"
}

@article{DElia:2011koc,
	author = "D'Elia, Massimo and Negro, Francesco",
	title = "{Chiral Properties of Strong Interactions in a Magnetic Background}",
	eprint = "1103.2080",
	archivePrefix = "arXiv",
	primaryClass = "hep-lat",
	doi = "10.1103/PhysRevD.83.114028",
	journal = "Phys. Rev. D",
	volume = "83",
	pages = "114028",
	year = "2011"
}

@article{Ilgenfritz:2012fw,
	author = "Ilgenfritz, E. -M. and Kalinowski, M. and Muller-Preussker, M. and Petersson, B. and Schreiber, A.",
	title = "{Two-color QCD with staggered fermions at finite temperature under the influence of a magnetic field}",
	eprint = "1203.3360",
	archivePrefix = "arXiv",
	primaryClass = "hep-lat",
	reportNumber = "HU-EP-12-09",
	doi = "10.1103/PhysRevD.85.114504",
	journal = "Phys. Rev. D",
	volume = "85",
	pages = "114504",
	year = "2012"
}

@article{Shushpanov:1997sf,
	author = "Shushpanov, I. A. and Smilga, Andrei V.",
	title = "{Quark condensate in a magnetic field}",
	eprint = "hep-ph/9703201",
	archivePrefix = "arXiv",
	reportNumber = "ITEP-TH-6-97, TPI-MINN-97-04, NUC-MINN-97-2-T",
	doi = "10.1016/S0370-2693(97)00441-3",
	journal = "Phys. Lett. B",
	volume = "402",
	pages = "351--358",
	year = "1997"
}

@article{Agasian:1999sx,
	author = "Agasian, Nikita O. and Shushpanov, I. A.",
	title = "{The Quark and gluon condensates and low-energy QCD theorems in a magnetic field}",
	eprint = "hep-ph/9911254",
	archivePrefix = "arXiv",
	doi = "10.1016/S0370-2693(99)01414-8",
	journal = "Phys. Lett. B",
	volume = "472",
	pages = "143--149",
	year = "2000"
}

@article{Alexandre:2000yf,
	author = "Alexandre, J. and Farakos, K. and Koutsoumbas, G.",
	title = "{Magnetic catalysis in QED(3) at finite temperature: Beyond the constant mass approximation}",
	eprint = "hep-th/0010211",
	archivePrefix = "arXiv",
	reportNumber = "NTUA-101-00",
	doi = "10.1103/PhysRevD.63.065015",
	journal = "Phys. Rev. D",
	volume = "63",
	pages = "065015",
	year = "2001"
}

@article{Agasian:2001hv,
	author = "Agasian, Nikita O.",
	title = "{Chiral thermodynamics in a magnetic field}",
	eprint = "hep-ph/0112341",
	archivePrefix = "arXiv",
	doi = "10.1134/1.1358481",
	journal = "Phys. Atom. Nucl.",
	volume = "64",
	pages = "554--560",
	year = "2001"
}

@article{Cohen:2007bt,
	author = "Cohen, Thomas D. and McGady, David A. and Werbos, Elizabeth S.",
	title = "{The Chiral condensate in a constant electromagnetic field}",
	eprint = "0706.3208",
	archivePrefix = "arXiv",
	primaryClass = "hep-ph",
	doi = "10.1103/PhysRevC.76.055201",
	journal = "Phys. Rev. C",
	volume = "76",
	pages = "055201",
	year = "2007"
}

@article{Gatto:2010qs,
	author = "Gatto, Raoul and Ruggieri, Marco",
	title = "{Dressed Polyakov loop and phase diagram of hot quark matter under magnetic field}",
	eprint = "1007.0790",
	archivePrefix = "arXiv",
	primaryClass = "hep-ph",
	reportNumber = "YITP-10-55",
	doi = "10.1103/PhysRevD.82.054027",
	journal = "Phys. Rev. D",
	volume = "82",
	pages = "054027",
	year = "2010"
}

@article{Gatto:2010pt,
	author = "Gatto, Raoul and Ruggieri, Marco",
	title = "{Deconfinement and Chiral Symmetry Restoration in a Strong Magnetic Background}",
	eprint = "1012.1291",
	archivePrefix = "arXiv",
	primaryClass = "hep-ph",
	reportNumber = "YITP-10-96",
	doi = "10.1103/PhysRevD.83.034016",
	journal = "Phys. Rev. D",
	volume = "83",
	pages = "034016",
	year = "2011"
}

@article{Mizher:2010zb,
	author = "Mizher, Ana Julia and Chernodub, M. N. and Fraga, Eduardo S.",
	title = "{Phase diagram of hot QCD in an external magnetic field: possible splitting of deconfinement and chiral transitions}",
	eprint = "1004.2712",
	archivePrefix = "arXiv",
	primaryClass = "hep-ph",
	doi = "10.1103/PhysRevD.82.105016",
	journal = "Phys. Rev. D",
	volume = "82",
	pages = "105016",
	year = "2010"
}

@article{Kashiwa:2011js,
	author = "Kashiwa, Kouji",
	title = "{Entanglement between chiral and deconfinement transitions under strong uniform magnetic background field}",
	eprint = "1104.5167",
	archivePrefix = "arXiv",
	primaryClass = "hep-ph",
	doi = "10.1103/PhysRevD.83.117901",
	journal = "Phys. Rev. D",
	volume = "83",
	pages = "117901",
	year = "2011"
}

@article{Avancini:2012ee,
	author = "Avancini, Sidney S. and Menezes, Debora P. and Pinto, Marcus B. and Providencia, Constanca",
	title = "{The QCD Critical End Point Under Strong Magnetic Fields}",
	eprint = "1202.5641",
	archivePrefix = "arXiv",
	primaryClass = "hep-ph",
	doi = "10.1103/PhysRevD.85.091901",
	journal = "Phys. Rev. D",
	volume = "85",
	pages = "091901",
	year = "2012"
}

@article{Andersen:2012dz,
	author = "Andersen, Jens O.",
	title = "{Thermal pions in a magnetic background}",
	eprint = "1202.2051",
	archivePrefix = "arXiv",
	primaryClass = "hep-ph",
	doi = "10.1103/PhysRevD.86.025020",
	journal = "Phys. Rev. D",
	volume = "86",
	pages = "025020",
	year = "2012"
}

@article{Scherer:2012nn,
	author = "Scherer, Daniel D. and Gies, Holger",
	title = "{Renormalization Group Study of Magnetic Catalysis in the 3d Gross-Neveu Model}",
	eprint = "1201.3746",
	archivePrefix = "arXiv",
	primaryClass = "cond-mat.str-el",
	doi = "10.1103/PhysRevB.85.195417",
	journal = "Phys. Rev. B",
	volume = "85",
	pages = "195417",
	year = "2012"
}

@article{Fukushima:2012kc,
	author = "Fukushima, Kenji and Hidaka, Yoshimasa",
	title = "{Magnetic Catalysis Versus Magnetic Inhibition}",
	eprint = "1209.1319",
	archivePrefix = "arXiv",
	primaryClass = "hep-ph",
	doi = "10.1103/PhysRevLett.110.031601",
	journal = "Phys. Rev. Lett.",
	volume = "110",
	number = "3",
	pages = "031601",
	year = "2013"
}

@article{Kojo:2012js,
	author = "Kojo, Toru and Su, Nan",
	title = "{The quark mass gap in a magnetic field}",
	eprint = "1211.7318",
	archivePrefix = "arXiv",
	primaryClass = "hep-ph",
	reportNumber = "BI-TP-2012-50",
	doi = "10.1016/j.physletb.2013.02.024",
	journal = "Phys. Lett. B",
	volume = "720",
	pages = "192--197",
	year = "2013"
}

@article{Bruckmann:2013oba,
	author = "Bruckmann, Falk and Endrodi, Gergely and Kovacs, Tamas G.",
	title = "{Inverse magnetic catalysis and the Polyakov loop}",
	eprint = "1303.3972",
	archivePrefix = "arXiv",
	primaryClass = "hep-lat",
	doi = "10.1007/JHEP04(2013)112",
	journal = "JHEP",
	volume = "04",
	pages = "112",
	year = "2013"
}

@article{Chao:2013qpa,
	author = "Chao, Jingyi and Chu, Pengcheng and Huang, Mei",
	title = "{Inverse magnetic catalysis induced by sphalerons}",
	eprint = "1305.1100",
	archivePrefix = "arXiv",
	primaryClass = "hep-ph",
	doi = "10.1103/PhysRevD.88.054009",
	journal = "Phys. Rev. D",
	volume = "88",
	pages = "054009",
	year = "2013"
}

@article{Fraga:2013ova,
	author = "Fraga, E. S. and Mintz, B. W. and Schaffner-Bielich, J.",
	title = "{A search for inverse magnetic catalysis in thermal quark-meson models}",
	eprint = "1311.3964",
	archivePrefix = "arXiv",
	primaryClass = "hep-ph",
	doi = "10.1016/j.physletb.2014.02.028",
	journal = "Phys. Lett. B",
	volume = "731",
	pages = "154--158",
	year = "2014"
}

@article{Ferreira:2014kpa,
	author = "Ferreira, M. and Costa, P. and Louren\c{c}o, O. and Frederico, T. and Provid\^encia, C.",
	title = "{Inverse magnetic catalysis in the (2+1)-flavor Nambu-Jona-Lasinio and Polyakov-Nambu-Jona-Lasinio models}",
	eprint = "1404.5577",
	archivePrefix = "arXiv",
	primaryClass = "hep-ph",
	doi = "10.1103/PhysRevD.89.116011",
	journal = "Phys. Rev. D",
	volume = "89",
	number = "11",
	pages = "116011",
	year = "2014"
}

@article{Farias:2014eca,
	author = "Farias, R. L. S. and Gomes, K. P. and Krein, G. I. and Pinto, M. B.",
	title = "{Importance of asymptotic freedom for the pseudocritical temperature in magnetized quark matter}",
	eprint = "1404.3931",
	archivePrefix = "arXiv",
	primaryClass = "hep-ph",
	doi = "10.1103/PhysRevC.90.025203",
	journal = "Phys. Rev. C",
	volume = "90",
	number = "2",
	pages = "025203",
	year = "2014"
}

@article{Yu:2014sla,
	author = "Yu, Lang and Liu, Hao and Huang, Mei",
	title = "{Spontaneous generation of local CP violation and inverse magnetic catalysis}",
	eprint = "1404.6969",
	archivePrefix = "arXiv",
	primaryClass = "hep-ph",
	doi = "10.1103/PhysRevD.90.074009",
	journal = "Phys. Rev. D",
	volume = "90",
	number = "7",
	pages = "074009",
	year = "2014"
}

@article{Andersen:2014oaa,
	author = "Andersen, Jens O. and Naylor, William R. and Tranberg, Anders",
	title = "{Inverse magnetic catalysis and regularization in the quark-meson model}",
	eprint = "1410.5247",
	archivePrefix = "arXiv",
	primaryClass = "hep-ph",
	doi = "10.1007/JHEP02(2015)042",
	journal = "JHEP",
	volume = "02",
	pages = "042",
	year = "2015"
}

@article{Ferrer:2014qka,
	author = "Ferrer, E. J. and de la Incera, V. and Wen, X. J.",
	title = "{Quark Antiscreening at Strong Magnetic Field and Inverse Magnetic Catalysis}",
	eprint = "1407.3503",
	archivePrefix = "arXiv",
	primaryClass = "nucl-th",
	doi = "10.1103/PhysRevD.91.054006",
	journal = "Phys. Rev. D",
	volume = "91",
	number = "5",
	pages = "054006",
	year = "2015"
}

@article{Providencia:2014txa,
	author = "Provid\^encia, Constan\c{c}a and Ferreira, M\'arcio and Costa, Pedro",
	title = "{Inverse magnetic catalysis in the Polyakov-Nambu-Jona-Lasinio and entangled Polyakov-Nambu-Jona-Lasinio models}",
	eprint = "1412.8308",
	archivePrefix = "arXiv",
	primaryClass = "hep-ph",
	doi = "10.5506/APhysPolBSupp.8.207",
	journal = "Acta Phys. Polon. Supp.",
	volume = "8",
	number = "1",
	pages = "207",
	year = "2015"
}

@article{Farias:2016gmy,
	author = "Farias, R. L. S. and Timoteo, V. S. and Avancini, S. S. and Pinto, M. B. and Krein, G.",
	title = "{Thermo-magnetic effects in quark matter: Nambu--Jona-Lasinio model constrained by lattice QCD}",
	eprint = "1603.03847",
	archivePrefix = "arXiv",
	primaryClass = "hep-ph",
	doi = "10.1140/epja/i2017-12320-8",
	journal = "Eur. Phys. J. A",
	volume = "53",
	number = "5",
	pages = "101",
	year = "2017"
}

@article{Mao:2016fha,
	author = "Mao, Shijun",
	title = "{Inverse magnetic catalysis in Nambu\textendash{}Jona-Lasinio model beyond mean field}",
	eprint = "1602.06503",
	archivePrefix = "arXiv",
	primaryClass = "hep-ph",
	doi = "10.1016/j.physletb.2016.05.018",
	journal = "Phys. Lett. B",
	volume = "758",
	pages = "195--199",
	year = "2016"
}

@article{Mamo:2015dea,
	author = "Mamo, Kiminad A.",
	title = "{Inverse magnetic catalysis in holographic models of QCD}",
	eprint = "1501.03262",
	archivePrefix = "arXiv",
	primaryClass = "hep-th",
	doi = "10.1007/JHEP05(2015)121",
	journal = "JHEP",
	volume = "05",
	pages = "121",
	year = "2015"
}

@article{Endrodi:2019whh,
	author = "Endr\H{o}di, Gergely and Mark\'o, Gergely",
	title = "{Magnetized baryons and the QCD phase diagram: NJL model meets the lattice}",
	eprint = "1905.02103",
	archivePrefix = "arXiv",
	primaryClass = "hep-lat",
	doi = "10.1007/JHEP08(2019)036",
	journal = "JHEP",
	volume = "08",
	pages = "036",
	year = "2019"
}

@article{Endrodi:2019zrl,
	author = "Endrodi, Gergely and Giordano, Matteo and Katz, Sandor D. and Kov\'acs, T. G. and Pittler, Ferenc",
	title = "{Magnetic catalysis and inverse catalysis for heavy pions}",
	eprint = "1904.10296",
	archivePrefix = "arXiv",
	primaryClass = "hep-lat",
	doi = "10.1007/JHEP07(2019)007",
	journal = "JHEP",
	volume = "07",
	pages = "007",
	year = "2019"
}

@article{Tomiya:2019nym,
	author = "Tomiya, Akio and Ding, Heng-Tong and Wang, Xiao-Dan and Zhang, Yu and Mukherjee, Swagato and Schmidt, Christian",
	title = "{Phase structure of three flavor QCD in external magnetic fields using HISQ fermions}",
	eprint = "1904.01276",
	archivePrefix = "arXiv",
	primaryClass = "hep-lat",
	doi = "10.22323/1.334.0163",
	journal = "PoS",
	volume = "LATTICE2018",
	pages = "163",
	year = "2019"
}

@article{Li:2016gfn,
    author = "Li, Danning and Huang, Mei and Yang, Yi and Yuan, Pei-Hung",
    title = "{Inverse Magnetic Catalysis in the Soft-Wall Model of AdS/QCD}",
    eprint = "1610.04618",
    archivePrefix = "arXiv",
    primaryClass = "hep-th",
    doi = "10.1007/JHEP02(2017)030",
    journal = "JHEP",
    volume = "02",
    pages = "030",
    year = "2017"
}

@article{Rodrigues:2018pep,
    author = "Rodrigues, Diego M. and Li, Danning and Folco Capossoli, Eduardo and Boschi-Filho, Henrique",
    title = "{Chiral symmetry breaking and restoration in 2+1 dimensions from holography: Magnetic and inverse magnetic catalysis}",
    eprint = "1807.11822",
    archivePrefix = "arXiv",
    primaryClass = "hep-th",
    doi = "10.1103/PhysRevD.98.106007",
    journal = "Phys. Rev. D",
    volume = "98",
    number = "10",
    pages = "106007",
    year = "2018"
}

@article{He:2020fdi,
	author = "He, Song and Yang, Yi and Yuan, Pei-Hung",
	title = "{Analytic Study of Magnetic Catalysis in Holographic QCD}",
	eprint = "2004.01965",
	archivePrefix = "arXiv",
	primaryClass = "hep-th",
	month = "4",
	year = "2020"
}

@ARTICLE{Kharzeev:2007tn,
	author = "Kharzeev, D. and Zhitnitsky, A.",
	title = "{Charge separation induced by P-odd bubbles in QCD matter}",
	eprint = "0706.1026",
	archivePrefix = "arXiv",
	primaryClass = "hep-ph",
	reportNumber = "BNL-NT-07-24",
	doi = "10.1016/j.nuclphysa.2007.10.001",
	journal = "Nucl. Phys. A",
	volume = "797",
	pages = "67--79",
	year = "2007"
}

@ARTICLE{Okorokov:2009bf,
	author = "Okorokov, V.A.",
	title = "{Estimation of P-odd correlators in heavy ion collisions at RHIC energies 62.4-GeV - 200-GeV}",
	eprint = "0908.2522",
	archivePrefix = "arXiv",
	primaryClass = "nucl-th",
	month = "8",
	year = "2009"
}

@ARTICLE{Fukushima:2008xe,
	author = "Fukushima, Kenji and Kharzeev, Dmitri E. and Warringa, Harmen J.",
	title = "{The Chiral Magnetic Effect}",
	eprint = "0808.3382",
	archivePrefix = "arXiv",
	primaryClass = "hep-ph",
	doi = "10.1103/PhysRevD.78.074033",
	journal = "Phys. Rev. D",
	volume = "78",
	pages = "074033",
	year = "2008"
}

@ARTICLE{Klevansky:1989vi,
	author = "Klevansky, S.P. and Lemmer, Richard H.",
	title = "{Chiral symmetry restoration in the Nambu-Jona-Lasinio model with a constant electromagnetic field}",
	doi = "10.1103/PhysRevD.39.3478",
	journal = "Phys. Rev. D",
	volume = "39",
	pages = "3478--3489",
	year = "1989"
}

@ARTICLE{Klimenko:1990rh,
	author = "Klimenko, K.G.",
	title = "{Three-dimensional Gross-Neveu model in an external magnetic field}",
	reportNumber = "IFVE-90-189",
	doi = "10.1007/BF01015908",
	journal = "Theor. Math. Phys.",
	volume = "89",
	pages = "1161--1168",
	year = "1992"
}

@ARTICLE{Gusynin:1995nb,
	author = "Gusynin, V.P. and Miransky, V.A. and Shovkovy, I.A.",
	title = "{Dimensional reduction and catalysis of dynamical symmetry breaking by a magnetic field}",
	eprint = "hep-ph/9509320",
	archivePrefix = "arXiv",
	reportNumber = "UCLA-95-TEP-26",
	doi = "10.1016/0550-3213(96)00021-1",
	journal = "Nucl. Phys. B",
	volume = "462",
	pages = "249--290",
	year = "1996"
}

@ARTICLE{Bali:2011qj,
	author = "Bali, G.S. and Bruckmann, F. and Endrodi, G. and Fodor, Z. and Katz, S.D. and Krieg, S. and Schafer, A. and Szabo, K.K.",
	title = "{The QCD phase diagram for external magnetic fields}",
	eprint = "1111.4956",
	archivePrefix = "arXiv",
	primaryClass = "hep-lat",
	doi = "10.1007/JHEP02(2012)044",
	journal = "JHEP",
	volume = "02",
	pages = "044",
	year = "2012"
}

@ARTICLE{Bali:2012zg,
	author = "Bali, G.S. and Bruckmann, F. and Endrodi, G. and Fodor, Z. and Katz, S.D. and Schafer, A.",
	title = "{QCD quark condensate in external magnetic fields}",
	eprint = "1206.4205",
	archivePrefix = "arXiv",
	primaryClass = "hep-lat",
	doi = "10.1103/PhysRevD.86.071502",
	journal = "Phys. Rev. D",
	volume = "86",
	pages = "071502",
	year = "2012"
}

@ARTICLE{Chernodub:2010qx,
	author = "Chernodub, M.N.",
	title = "{Superconductivity of QCD vacuum in strong magnetic field}",
	eprint = "1008.1055",
	archivePrefix = "arXiv",
	primaryClass = "hep-ph",
	doi = "10.1103/PhysRevD.82.085011",
	journal = "Phys. Rev. D",
	volume = "82",
	pages = "085011",
	year = "2010"
}

@ARTICLE{Chernodub:2011mc,
	author = "Chernodub, M.N.",
	title = "{Spontaneous electromagnetic superconductivity of vacuum in strong magnetic field: evidence from the Nambu--Jona-Lasinio model}",
	eprint = "1101.0117",
	archivePrefix = "arXiv",
	primaryClass = "hep-ph",
	doi = "10.1103/PhysRevLett.106.142003",
	journal = "Phys. Rev. Lett.",
	volume = "106",
	pages = "142003",
	year = "2011"
}

@ARTICLE{Klevansky:1992qe,
	author = "Klevansky, S. P.",
	title  = "{The Nambu\textendash Jona-Lasinio model of quantum chromodynamics}",
	doi     = "10.1103/RevModPhys.64.649",
	journal = "Rev. Mod. Phys.",
	volume  = "64",
	year    = "1992",
	pages   = "649-708",
}

@ARTICLE{Fayazbakhsh:2012vr,
	author = "Fayazbakhsh, Sh. and Sadeghian, S. and Sadooghi, N.",
	title = "{Properties of neutral mesons in a hot and magnetized quark matter}",
	eprint = "1206.6051",
	archivePrefix = "arXiv",
	primaryClass = "hep-ph",
	doi = "10.1103/PhysRevD.86.085042",
	journal = "Phys. Rev. D",
	volume = "86",
	pages = "085042",
	year = "2012"
}

@ARTICLE{Fayazbakhsh:2013cha,
	author = "Fayazbakhsh, Sh. and Sadooghi, N.",
	title = "{Weak decay constant of neutral pions in a hot and magnetized quark matter}",
	eprint = "1306.2098",
	archivePrefix = "arXiv",
	primaryClass = "hep-ph",
	doi = "10.1103/PhysRevD.88.065030",
	journal = "Phys. Rev. D",
	volume = "88",
	number = "6",
	pages = "065030",
	year = "2013"
}

@ARTICLE{Wang:2017vtn,
	author = "Wang, Ziyue and Zhuang, Pengfei",
	title = "{Meson properties in magnetized quark matter}",
	eprint = "1712.00554",
	archivePrefix = "arXiv",
	primaryClass = "hep-ph",
	doi = "10.1103/PhysRevD.97.034026",
	journal = "Phys. Rev. D",
	volume = "97",
	number = "3",
	pages = "034026",
	year = "2018"
}

@article{Sheng:2020hge,
	author = "Sheng, Bingkai and Wang, Yuanyuan and Wang, Xinyang and Yu, Lang",
	title = "{Pole and screening masses of neutral pions in a hot and magnetized medium: A comprehensive study in the Nambu\textendash{}Jona-Lasinio model}",
	eprint = "2010.05716",
	archivePrefix = "arXiv",
	primaryClass = "hep-ph",
	doi = "10.1103/PhysRevD.103.094001",
	journal = "Phys. Rev. D",
	volume = "103",
	number = "9",
	pages = "094001",
	year = "2021"
}

@ARTICLE{Cheng:2010fe,
	author = "Cheng, M. and others",
	title = "{Meson screening masses from lattice QCD with two light and the strange quark}",
	eprint = "1010.1216",
	archivePrefix = "arXiv",
	primaryClass = "hep-lat",
	doi = "10.1140/epjc/s10052-011-1564-y",
	journal = "Eur. Phys. J. C",
	volume = "71",
	pages = "1564",
	year = "2011"
}

@article{Maezawa:2013nxa,
	author = "Maezawa, Yu and Bazavov, Alexei and Karsch, Frithjof and Petreczky, Peter and Mukherjee, Swagato",
	title = "{Meson screening masses at finite temperature with Highly Improved Staggered Quarks}",
	eprint = "1312.4375",
	archivePrefix = "arXiv",
	primaryClass = "hep-lat",
	doi = "10.22323/1.187.0149",
	journal = "PoS",
	volume = "LATTICE2013",
	pages = "149",
	year = "2014"
}

@article{Kaczmarek:2013kva,
	author = {Kaczmarek, Olaf and Laermann, Edwin and M\"uller, Marcel},
	title = "{The thermodynamic and the continuum limit of meson screening masses}",
	eprint = "1311.3889",
	archivePrefix = "arXiv",
	primaryClass = "hep-lat",
	doi = "10.22323/1.187.0150",
	journal = "PoS",
	volume = "LATTICE2013",
	pages = "150",
	year = "2014"
}

@ARTICLE{Bazavov:2019www,
	author = "Bazavov, Alexei and others",
	title = "{Meson screening masses in (2+1)-flavor QCD}",
	eprint = "1908.09552",
	archivePrefix = "arXiv",
	primaryClass = "hep-lat",
	doi = "10.1103/PhysRevD.100.094510",
	journal = "Phys. Rev. D",
	volume = "100",
	number = "9",
	pages = "094510",
	year = "2019"
}

@article{Cao:2021tcr,
author = "Cao, Xuanmin and Qiu, Songyu and Liu, Hui and Li, Danning",
title = "{Thermal properties of light mesons from holography}",
eprint = "2102.10946",
archivePrefix = "arXiv",
primaryClass = "hep-ph",
doi = "10.1007/JHEP08(2021)005",
journal = "JHEP",
volume = "08",
pages = "005",
year = "2021"
}

@article{Kunihiro:1991hp,
author = "Kunihiro, T.",
title = "{Chiral restoration, flavor symmetry and the axial anomaly at finite temperature in an effective theory}",
doi = "10.1016/S0550-3213(05)80035-5",
journal = "Nucl. Phys. B",
volume = "351",
pages = "593--622",
year = "1991"
}

@article{Florkowski:1993br,
	author = "Florkowski, Wojciech and Friman, Bengt L.",
	title = "{Screening of the meson fields in the Nambu-Jona-Lasinio model}",
	reportNumber = "INP-1622-PH",
	journal = "Acta Phys. Polon. B",
	volume = "25",
	pages = "49--71",
	year = "1994"
}

@article{Wang:2013wk,
author = "Wang, Kun-lun and Liu, Yu-xin and Chang, Lei and Roberts, Craig D. and Schmidt, Sebastian M.",
title = "{Baryon and meson screening masses}",
eprint = "1301.6762",
archivePrefix = "arXiv",
primaryClass = "nucl-th",
doi = "10.1103/PhysRevD.87.074038",
journal = "Phys. Rev. D",
volume = "87",
number = "7",
pages = "074038",
year = "2013"
}

@ARTICLE{PhysRev.82.664,
	title = {On Gauge Invariance and Vacuum Polarization},
	author = {Schwinger, Julian},
	journal = {Phys. Rev.},
	volume = {82},
	issue = {5},
	pages = {664--679},
	numpages = {0},
	year = {1951},
	month = {Jun},
	publisher = {American Physical Society},
	doi = {10.1103/PhysRev.82.664},
	url = {https://link.aps.org/doi/10.1103/PhysRev.82.664}
}

@ARTICLE{Florkowski:1997pi,
	author = "Florkowski, Wojciech",
	title = "{Description of hot compressed hadronic matter based on an effective chiral Lagrangian}",
	eprint = "hep-ph/9701223",
	archivePrefix = "arXiv",
	reportNumber = "INP-1739-PH",
	journal = "Acta Phys. Polon. B",
	volume = "28",
	pages = "2079--2205",
	year = "1997"
}

@BOOK{book:63138,
	title =     {Table of integrals, series, and products},
	author =    {I.S. Gradshteyn and I.M. Ryzhik},
	publisher = {Academic Press},
	isbn =      {0123736374,0080471110,9780123738622,9780123736376,9780080471112,0123738628},
	year =      {2007},
	series =    {},
	edition =   {7th ed},
	volume =    {},
	url =       {http://gen.lib.rus.ec/book/index.php?md5=7116fc0ce3bbd85b1d2b5fd4e21beb10}
}

@ARTICLE{Ishii:2013kaa,
	author = "Ishii, Masahiro and Sasaki, Takahiro and Kashiwa, Kouji and Kouno, Hiroaki and Yahiro, Masanobu",
	title = "{Effective model approach to meson screening masses at finite temperature}",
	eprint = "1312.7424",
	archivePrefix = "arXiv",
	primaryClass = "hep-ph",
	doi = "10.1103/PhysRevD.89.071901",
	journal = "Phys. Rev. D",
	volume = "89",
	number = "7",
	pages = "071901",
	year = "2014"
}

@ARTICLE{Ishii:2015ira,
	author = "Ishii, Masahiro and Yonemura, Koji and Takahashi, Junichi and Kouno, Hiroaki and Yahiro, Masanobu",
	title = "{Determination of $U(1)_{\rm A}$ restoration from pion and $a_0$-meson screening masses: Toward the chiral regime}",
	eprint = "1504.04463",
	archivePrefix = "arXiv",
	primaryClass = "hep-ph",
	doi = "10.1103/PhysRevD.93.016002",
	journal = "Phys. Rev. D",
	volume = "93",
	number = "1",
	pages = "016002",
	year = "2016"
}

@ARTICLE{Hellstern:1997nv,
	author = "Hellstern, G. and Alkofer, Reinhard and Reinhardt, H.",
	title = "{Diquark confinement in an extended NJL model}",
	eprint = "hep-ph/9706551",
	archivePrefix = "arXiv",
	reportNumber = "UNITU-THEP-11-97, TU-GK-5-97",
	doi = "10.1016/S0375-9474(97)00412-0",
	journal = "Nucl. Phys. A",
	volume = "625",
	pages = "697--712",
	year = "1997"
}

@ARTICLE{Ebert:1996vx,
	author = "Ebert, Dietmar and Feldmann, Thorsten and Reinhardt, Hugo",
	title = "{Extended NJL model for light and heavy mesons without q - anti-q thresholds}",
	eprint = "hep-ph/9608223",
	archivePrefix = "arXiv",
	reportNumber = "DESY-96-214, HUB-EP-96-40",
	doi = "10.1016/0370-2693(96)01158-6",
	journal = "Phys. Lett. B",
	volume = "388",
	pages = "154--160",
	year = "1996"
}

@ARTICLE{Bentz:2001vc,
	author = "Bentz, Wolfgang and Thomas, Anthony William",
	title = "{The Stability of nuclear matter in the Nambu-Jona-Lasinio model}",
	eprint = "nucl-th/0105022",
	archivePrefix = "arXiv",
	reportNumber = "ADP-01-09-T444",
	doi = "10.1016/S0375-9474(01)01119-8",
	journal = "Nucl. Phys. A",
	volume = "696",
	pages = "138--172",
	year = "2001"
}

@ARTICLE{Shovkovy:2012zn,
	author = "Shovkovy, Igor A.",
	title = "{Magnetic Catalysis: A Review}",
	eprint = "1207.5081",
	archivePrefix = "arXiv",
	primaryClass = "hep-ph",
	doi = "10.1007/978-3-642-37305-3_2",
	journal = "Lect. Notes Phys.",
	volume = "871",
	pages = "13--49",
	year = "2013"
}

@article{Sheng:2021evj,
    author = "Sheng, Bing-kai and Wang, Xinyang and Yu, Lang",
    title = "{Impacts of inverse magnetic catalysis on screening masses of neutral pions and sigma mesons in hot and magnetized quark matter}",
    eprint = "2110.12811",
    archivePrefix = "arXiv",
    primaryClass = "hep-ph",
    doi = "10.1103/PhysRevD.105.034003",
    journal = "Phys. Rev. D",
    volume = "105",
    number = "3",
    pages = "034003",
    year = "2022"
}

@article{Ding:2025pbu,
    author = "Ding, Heng-Tong and Gu, Jin-Biao and Li, Sheng-Tai and Thakkar, Rishabh",
    title = "{Chiral condensates and screening masses of neutral pseudoscalar mesons from lattice QCD at physical quark masses}",
    eprint = "2501.11262",
    archivePrefix = "arXiv",
    primaryClass = "hep-lat",
    doi = "10.1103/PhysRevD.111.074513",
    journal = "Phys. Rev. D",
    volume = "111",
    number = "7",
    pages = "074513",
    year = "2025"
}

@article{Ding:2022tqn,
    author = "Ding, H. -T. and Li, S. -T. and Liu, J. -H. and Wang, X. -D.",
    title = "{Chiral condensates and screening masses of neutral pseudoscalar mesons in thermomagnetic QCD medium}",
    eprint = "2201.02349",
    archivePrefix = "arXiv",
    primaryClass = "hep-lat",
    doi = "10.1103/PhysRevD.105.034514",
    journal = "Phys. Rev. D",
    volume = "105",
    number = "3",
    pages = "034514",
    year = "2022"
}

@article{Coppola:2024uvz,
    author = "Coppola, M{\'a}ximo and Tavares, William R. and Avancini, Sidney S. and Sodr{\'e}, Joana C. and Scoccola, Norberto N.",
    title = "{Thermomagnetic effects on light pseudoscalar meson masses within the SU(3) Nambu{\textendash}Jona-Lasinio model}",
    eprint = "2410.05568",
    archivePrefix = "arXiv",
    primaryClass = "hep-ph",
    doi = "10.1103/PhysRevD.110.114036",
    journal = "Phys. Rev. D",
    volume = "110",
    number = "11",
    pages = "114036",
    year = "2024"
}

@article{Wang:2021dcy,
    author = "Wang, Yuanyuan and Matsuzaki, Shinya",
    title = "{Axial inverse magnetic catalysis}",
    eprint = "2110.10432",
    archivePrefix = "arXiv",
    primaryClass = "hep-ph",
    doi = "10.1103/PhysRevD.105.074015",
    journal = "Phys. Rev. D",
    volume = "105",
    number = "7",
    pages = "074015",
    year = "2022"
}

@article{Mei:2020jzn,
    author = "Mei, Jie and Mao, Shijun",
    title = "{Inverse catalysis effect of the quark anomalous magnetic moment to chiral restoration and deconfinement phase transitions}",
    eprint = "2008.12123",
    archivePrefix = "arXiv",
    primaryClass = "hep-ph",
    doi = "10.1103/PhysRevD.102.114035",
    journal = "Phys. Rev. D",
    volume = "102",
    number = "11",
    pages = "114035",
    year = "2020"
}

@article{Ding:2026ewc,
    author = "Ding, Heng-Tong and Hern{\'a}ndez Hern{\'a}ndez, Jos{\'e} Javier and Zhang, Dan",
    title = "{Chiral and $U(1)_A$ symmetries in background magnetic fields from lattice QCD}",
    eprint = "2607.11625",
    archivePrefix = "arXiv",
    primaryClass = "hep-lat",
    month = "7",
    year = "2026"
}

@misc{Yu:inpreparation,
  author = {Yu, Lang},
    note   = {Manuscript in preparation}
}

@article{Buchoff:2013nra,
    author = "Buchoff, Michael I. and others",
    title = "{QCD chiral transition, U(1)A symmetry and the dirac spectrum using domain wall fermions}",
    eprint = "1309.4149",
    archivePrefix = "arXiv",
    primaryClass = "hep-lat",
    doi = "10.1103/PhysRevD.89.054514",
    journal = "Phys. Rev. D",
    volume = "89",
    number = "5",
    pages = "054514",
    year = "2014"
}

@article{Moreira:2020wau,
    author = "Moreira, Jo{\~a}o and Costa, Pedro and Restrepo, Tulio E.",
    title = "{Magnetic field dependent {\textquoteright}t Hooft determinant extended Nambu{\textendash}Jona-Lasinio model}",
    eprint = "2005.07049",
    archivePrefix = "arXiv",
    primaryClass = "hep-ph",
    doi = "10.1103/PhysRevD.102.014032",
    journal = "Phys. Rev. D",
    volume = "102",
    number = "1",
    pages = "014032",
    year = "2020"
}

@article{Bhattacharya:2014ara,
    author = "Bhattacharya, Tanmoy and others",
    title = "{QCD Phase Transition with Chiral Quarks and Physical Quark Masses}",
    eprint = "1402.5175",
    archivePrefix = "arXiv",
    primaryClass = "hep-lat",
    reportNumber = "BNL-103837-2014-JA, CU-TP-1205, INT-PUB-14-003, LLNL-JRNL-650194",
    doi = "10.1103/PhysRevLett.113.082001",
    journal = "Phys. Rev. Lett.",
    volume = "113",
    number = "8",
    pages = "082001",
    year = "2014"
}

@article{Zhai:2026cud,
    author = "Zhai, Shijie and Mei, Jie and Yu, Lang and Huang, Mei",
    title = "{Magnetized QCD Matter}",
    doi = "10.3390/universe12060154",
    journal = "Universe",
    volume = "12",
    number = "6",
    pages = "154",
    year = "2026"
}

\end{document}